\documentclass[showpacs, pre]{revtex4}   %single column 2 space for revision process

\usepackage{amsmath}
\usepackage{amssymb}
\usepackage{times}
\usepackage{verbatim}
\usepackage[applemac]{inputenc}
\usepackage[T1]{fontenc}
\usepackage{lmodern}
\usepackage{wrapfig}
\usepackage{graphicx}
\usepackage{rotating}
\usepackage{multirow}
\usepackage{float}

\DeclareGraphicsExtensions{.eps}
\graphicspath{{}}

\begin{document}

\title{Evolution of Cooperation on Spatially Embedded Networks}
\author{Pierre Buesser}
\email{pierre.buesser@unil.ch}
\affiliation{Information Systems Institute, HEC, University of Lausanne, Switzerland}

\author{Marco Tomassini}
\email{marco.tomassini@unil.ch}
\affiliation{Information Systems Institute, HEC, University of Lausanne, Switzerland}

\begin{abstract}

In this work we study the behavior of classical two-person, two-stra\-te\-gies evolutionary games
on networks embedded in a Euclidean two-dimensional space with different kinds of degree distributions and 
topologies going from regular to random, and to scale-free
ones. Using several imitative microscopic dynamics, 
we study the evolution of global cooperation on the above network classes and find that specific topologies having a hierarchical structure and 
an inhomogeneous degree distribution, such as Apollonian and grid-based  networks, are very conducive to cooperation.
Spatial scale-free networks are still good for cooperation but to a lesser degree. Both classes of networks enhance average cooperation in
all games with respect to standard random geometric graphs and regular grids by shifting the boundaries between cooperative and
defective regions.
These findings might be useful in the design of interaction structures that maintain cooperation when the agents are constrained to live 
in physical two-dimensional space.
%We check our results with several update rules.
\end{abstract}


\maketitle

\section{Introduction}
\label{abstract}

In a strategic context, Game Theory  is an ''ìnteractive decision theory`` where an agent's 
optimal action for herself depends on expectations on the actions of other agents - including herself~\cite{myerson}.
This
approach has proved very useful in a number of settings in biology, economy, and social science. 
Evolutionary
game theory in particular is well suited to the study of strategic interactions in animal and human populations that are large and
well-mixed in the sense that any agent can interact with any other agent in the population
(see e.g.~\cite{weibull95,Hofbauer1998,vega-redondo-03}). However, lattice-structured populations were used starting with 
the works of Axelrod on the repeated
Prisoner's Dilemma~\cite{axe84} and of Nowak and May on the one-shot case~\cite{nowakmay92}.
Especially in the last few years, population structures with local interactions have been brought to the focus
of research as it is known that social interactions can be better represented
by networks of contacts in which nodes represent agents and links stand for their 
relationships. This literature is already rather abundant and  it is difficult to be exhaustive;
good recent reviews can be found in~\cite{szabo,anxo1,mini-rev} and important foundational work dealing with the microscopic
agent dynamics appears, among others, in~\cite{traulsen2007pairwise,altrock2009fixation,wu2010universality,ohtsuki2006simple}. 
Most of the recent work has dealt with populations of agents
structured as non-spatial graphs (relational graphs), i.e. networks in which there is no underlying spatial structure and distances
are measured in terms of edge hops. Relational networks are adequate in many cases; for instance, when two people have
a connection in Facebook, for the purposes of the electronic communication, their actual physical distance is irrelevant, although
many links in the network will be related to closeness in space.
 However, often it is the case that actual distances matter; for example,
networks such as the road or the railway networks are of this type. Thus, while the recent  focus in complex network research 
has been mainly on relational
graphs, spatial graphs are also very important and have attracted attention (see~\cite{Barthelemy} for an excellent recent review).

In evolutionary games regular graphs such
as one- and two-dimensional lattices  have been used early on to provide a local structure to the population of 
interacting agents~\cite{axe84,nowakmay92}. These
networks can be considered relational for certain purposes but can also be trivially embedded in some low-dimensional
Euclidean space, with the associated distance metric. Practically the totality of the work on spatial evolutionary games has been done
on this kind of structure and a large literature has been produced (for a summary with references to 
previous work see~\cite{nowak-sig-00}). To the best of our knowledge, only few works have dealt  with spatial
networks other than grids in games e.g.~\cite{NamingGame,ApolloniusCoop2011}. In~\cite{NamingGame} 
geometric random graphs, which are Euclidean graphs built by drawing links between nodes that are
within a given distance, and spatially
embedded Watts--Strogatz networks are used in connection with the Naming Game. Ref.~\cite{ApolloniusCoop2011} 
deals with games on Apollonian networks, which can be viewed as Euclidean networks that can be built by recursively joining a new 
node in the interior of a triangle
with the nodes at its vertices. This kind of graphs  will be referred to later in the present work.
Clearly,  grids are only an approximation to actual spatial graphs representing
networks of contacts  in the physical world~\cite{Barthelemy}.

In the present work we extend the ideas and methods of evolutionary game theory to fixed spatial networks that go
beyond the much-studied discrete two-dimensional  lattices.
Previous results show that networks with heterogeneous degree distributions increase cooperation in the Hawks-Dowes games, while regular lattices increase cooperation in the Stag-Hunt games~\cite{santos-pach-05,santos-pach-06,anxo1}. Therefore we study spatial network models with a scale-free degree distribution as a first step. In a second part, by extending a previous work~\cite{ApolloniusCoop2011}, we show that Apollonian networks~\cite{Apollonius2005Herrmann} are such that both the benefits of spatiality and scale-free degree distribution can be gathered.  Some previous works use reduced game spaces. As these settings are not suitable for our discussion, we extend them to a larger game space.   Mobility (see e.g.~\cite{Meloni,Helb-Mobil}) is an important issue in these spatial networks but, as an obvious first step, here we shall deal only with static networks. Our exploratory approach is  based on numerical Monte Carlo simulations since an exact analytical description is essentially only possible in the mean-field case.

\section{Evolutionary Games on Networks}

\subsection{The Games Studied}

We investigate three classical two-person, two-strategy, symmetric games classes, namely the Prisoner's
Dilemma (PD),
the Hawk-Dove Game (HD), and the Stag Hunt (SH). These three games are simple metaphors for different kinds
of dilemmas that arise when individual and social interests collide. The Harmony game (H) is included for completeness but it
doesn't originate any conflict. 
The main features of these games are summarized here for completeness; more detailed accounts
can be found elsewhere e.g.~\cite{weibull95,Hofbauer1998,vega-redondo-03}.
The games have the generic payoff matrix $M$ (equation~\ref{eq:payoff}) which refers to the payoffs of the row player. The payoff matrix for the column player
is simply the transpose $M^\top$ since the game is symmetric.

\vspace{-0.5cm}
\begin{equation}
	\bordermatrix{\text{}&C& D\cr
	C&R&S\cr
	D&T&P\cr
	}
\label{eq:payoff}
\end{equation}
\vspace{0.05cm}
\noindent The set of strategies is $\Lambda=\{C,D\}$, where $C$ stands for ``cooperation'' and $D$ means ``defection''.
In the payoff matrix $R$ stands for the \textit{reward}
the two players receive if they
both cooperate, $P$ is the \textit{punishment} if they both defect, and $T$  is the
\textit{temptation}, i.e.~the payoff that a player receives if he defects while the
other cooperates getting the \textit{sucker's payoff} $S$.
In order to study the usual standard parameter space~\cite{santos-pach-06,anxo1}, we restrict the payoff values in the following
way: $R=1$, $P=0$, $-1 \leq S \leq 1$, and $0 \leq T \leq 2$. 

\noindent For the PD, the payoff values are ordered such that $T > R > P > S$. 
Defection is always the best rational individual choice, so that 
$(D,D)$ is the unique Nash Equilibrium (NE) and also the only fixed point of the replicator dynamics~\cite{weibull95,Hofbauer1998}.
Mutual cooperation  would be socially preferable but $C$ is strongly dominated by $D$. 

\noindent In the HD game, the order of $P$ and $S$ is reversed, yielding $T > R > S > P$. Thus,
when both players defect they each get the lowest payoff. 
Players have  a strong incentive
to play $D$, which is harmful for both parties if the outcome produced happens to be $(D,D)$.
$(C,D)$ and $(D,C)$ are NE of the game in pure strategies. There is
a third equilibrium in mixed strategies which is the only dynamically stable equilibrium~\cite{weibull95,Hofbauer1998}.

\noindent In the SH game, the ordering is $R > T > P > S$, which means that mutual cooperation $(C,C)$ is the best outcome,
Pareto-superior, and a NE.  Pareto-superior means that the equilibrium is a set of strategies, one for each player, such that there is 
no other strategy profile in which all players receive payoffs at least as high, and at least one player receives a strictly higher 
payoff.
 The second NE, where both players defect
is less efficient but also less risky. The tension is represented by the fact that the
socially preferable coordinated equilibrium $(C,C)$ might be missed for ``fear'' that the other player
will play $D$ instead.  The third mixed-strategy NE in the game is evolutionarily unstable~\cite{weibull95,Hofbauer1998}.

\noindent Finally, in the H game $R>S>T>P$ or $R>T>S>P $. In this case $C$ strongly dominates $D$ and
the trivial unique NE is $(C,C)$. The game is non-conflictual by definition and does not cause any
dilemma, it is mentioned to complete the quadrants of the parameter space.

\noindent In the $TS$-plane each game class corresponds
to a different quadrant depending on the above ordering of the payoffs as depicted in Fig~\ref{complet}, left image, and the
figures that follow.
We finally remark that a rigorous study of the evolutionary dynamics of $2 \times 2$ matrix games in finite mixing populations
has been published by Antal and Scheuring~\cite{Antal06}.

\subsection{Population Structure}
\label{pop-str}

The population of players is a connected undirected  graph $G(V,E)$, where the
 set of vertices $V$ represents the agents, while the set of edges  $E$ represents their symmetric interactions. The
 population size $N$ is the cardinality of $V$. The set of neighbors $V_i $ of an agent $i$ are the agents that are
 directly connected to $i$; the cardinality $|V_i|$ is the degree $k_i$ of vertex $i \in V$. The average
degree of the network is called $\langle k \rangle$, and $p(k)$ is the network's degree distribution function.

\subsection{Payoff Calculation and Strategy Update Rules}
\label{revision-protocols}

We need to specify how individual's payoffs are computed and how
agents decide to revise their current strategy, taking into account that each agent only interacts locally with its first
neighbors, not globally as in well mixed populations.
\noindent Let $\sigma_i(t)$ be a vector
giving the strategy profile at time $t$ with $C= (1, 0)$ and $D = (0, 1)$ and let $M$ be the payoff matrix of the game (equation~\ref{eq:payoff}). 
The quantity
\begin{equation}
\Pi_i(t) =  \sum _{j \in V_i} \sigma_i(t)\; M\; \sigma_{j}^\top(t)
\label{payoffs}
\end{equation}
is the cumulated payoff collected by player $i$ at time step $t$.

We use an asynchronous scheme for strategy update, i.e. players are updated one by one by choosing a random player in each step.  Several strategy update rules are customary in evolutionary game theory.
Here we shall describe the four imitative update protocols that have been used in our simulations. The first three  are well known;  
we thank an anonymous reviewer for suggesting a rule very similar to the fourth one presented here.

\noindent The \textit{local fitness-proportional} rule is stochastic and gives rise to replicator dynamics (RD)~\cite{Helbing92,hauer-doeb-2004}.  
Player $i$'s strategy $\sigma_i$ is updated by randomly drawing
another player $j$  from the neighborhood $V_i$,
and replacing $\sigma_i$ by $\sigma_j$ with probability: 

\begin{equation}
  p(\sigma_i \rightarrow \sigma_j)  = \left\{
  \begin{array}{l l}
     (\Pi_j - \Pi_i)/K & \quad \text{if $ \Pi_j > \Pi_i$}\\
    0 & \quad \text{if $\Pi_j \le \Pi_i$}\\
  \end{array} \right.
\label{proba}
\end{equation}
where $\Pi_j -\Pi_i$ is the difference of the payoffs earned by $j$ and $i$ respectively.
$K=\max(k_i,k_j)[(\max(1,T)-\min(0,S)]$ ensures
proper normalization of the probability $p(\sigma_i \rightarrow \sigma_j)$.
This normalization increase the frequency of imitations between nodes with smaller degree.
A more flexible update rule without the problem of normalization is the \textit{Fermi} rule. Here the randomly chosen player $i$ is given the
 opportunity to imitate a randomly chosen neighbor $j$ with probability :
\begin{equation}
  p(\sigma_i \rightarrow \sigma_j)  = \frac{1}{ 1+exp(-\beta(\Pi_j - \Pi_i))}
\label{fermi}
\end{equation}
where $\beta$ is a constant corresponding to the inverse temperature of the system, i.e. high temperature implies that imitation is random to a large extent and depends little on the payoffs. Thus when $\beta \to 0$ the probability of imitating $j$ tends to a constant value $0.5$ and when $\beta \to \infty$ the rule becomes deterministic: $i$ imitates $j$ if  $(\Pi_j - \Pi_i)>0$, otherwise it doesn't. For $\beta \subset [1.0,10.0]$ the rule leads approximatively to similar results as the \textit{local fitness-proportional} one. 
\noindent Another imitative strategy update protocol
is to switch to the strategy of the neighbor that has scored
best in the last time step. In contrast with the previous one, this rule is deterministic.
This \textit{imitation of the best} (IB) policy can be described in the following way:
the strategy $s_i(t)$ of individual $i$ at time step $t$ will be
\begin{equation}
s_i(t) = s_j(t-1),
\label{ib}
\end{equation}
where
\begin{equation}
j \in \{V_i \cup i\} \;s.t.\; \Pi_j = \max \{\Pi_k(t-1)\}, \; \forall k \in \{V_i \cup i\}.
\label{ib2}
\end{equation}
\noindent That is, individual $i$ will adopt the strategy of the player with the highest
payoff among its neighbors including itself.
If there is a tie, the winner individual is chosen uniformly at random.
The next update rule is a randomized version of the imitation of the best that we call IBR. Here player $i$ imitates player $j$ according to formula~\ref{proba}, but K is such that $\sum_{j \in V_{i}} p_{ij}=1$. 

A final remark is in order here. The above model rules are common and almost standard in numerical simulation
work, which has the advantage that the mathematics is simpler and results can be compared with previous work such as, for instance,~\cite{santos-pach-06,anxo1}.
However, it is far from clear whether these rules are representative of the ways in which human players actually take
their strategic decisions, as has been shown by many laboratory experiments. In these experiments it seems that 
learning and heuristics play an important role. Moreover, players are inhomogeneous in their behavior while
our stereotyped automata all behave in the same way and never change or adapt. Some less conventional work along
these lines can be found in~\cite{cardillo2010co,szolnoki2011imitating}. In Cardillo et al.~\cite{cardillo2010co} 
standard strategy update rules are
used but they are permitted to co-evolve with the agent's strategies. In Szolnoki et al.~\cite{szolnoki2011imitating}, rather than imitating strategies, agents
imitate a proxy that stands for emotions among their neighbors.

 \subsection{Simulation Parameters}

 The $TS$-plane has been sampled with a grid step of $0.1$ and
 each value in the phase space reported in the figures is the average of $50$ independent runs using a fresh graph 
 realization for each run, except for strictly regular or degree-invariant networks.  
 The evolution proceeds by first initializing the players at the nodes of the network with one of the two strategies 
 uniformly at random
such that each strategy has a fraction of approximately $1/2$ unless otherwise stated.
For each grid point, agents in the population are chosen sequentially at random to revise their strategies (asynchronous updating). Payoffs are updated after each strategy change. We let the system evolve for a period of $2*N$  time steps.
 At this point the system has reached a
 steady state in which the frequency of cooperators is stable except for small fluctuations. We then let the system evolve for $300$ further
 steps and take the average cooperation value in this interval. We repeat the whole process $50$ times for each grid point
  and, finally, we report the average cooperation values over those $50$ repetitions.

\section{Results}

In the following sections we investigate how spatiality affects cooperation through its effect on network topology. In section~\ref{SFSN} we study a class of spatial scale-free networks, and in section~\ref{AN} we study Apollonian networks, a different model of spatial scale-free network leading to high levels of cooperation.
Spatial scale-free networks and Apollonian networks combine spatiality and scale-free degree distribution. In section~\ref{supergrid} we propose
 a class of hierarchical spatial networks derived from lattices and random geometric graphs which also improve cooperation.

 \begin{figure*}[ht!]
\begin{center}
\begin{tabular} {cccccccc} 
\includegraphics[width=6cm]{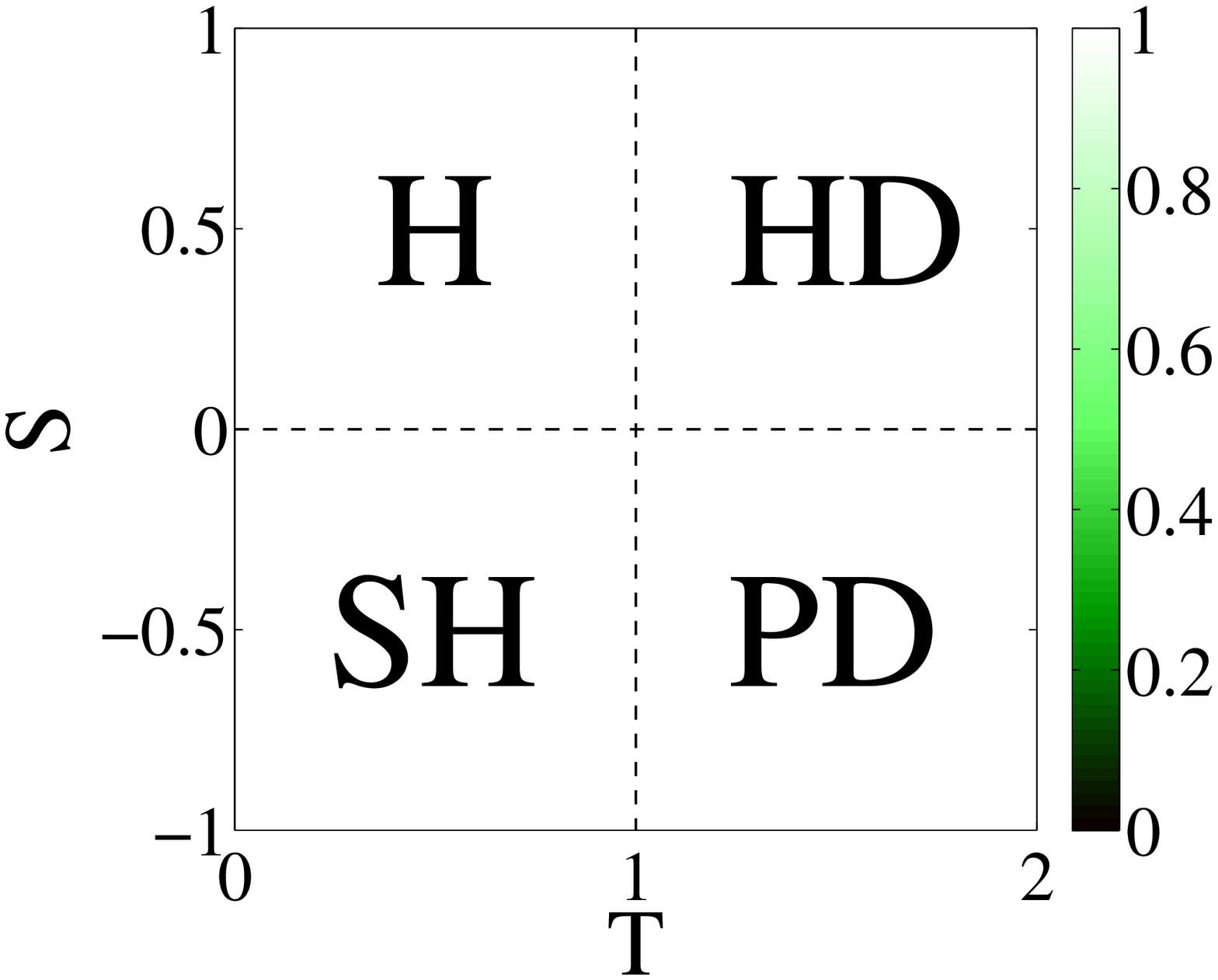} &%STPlane
\includegraphics[width=6cm]{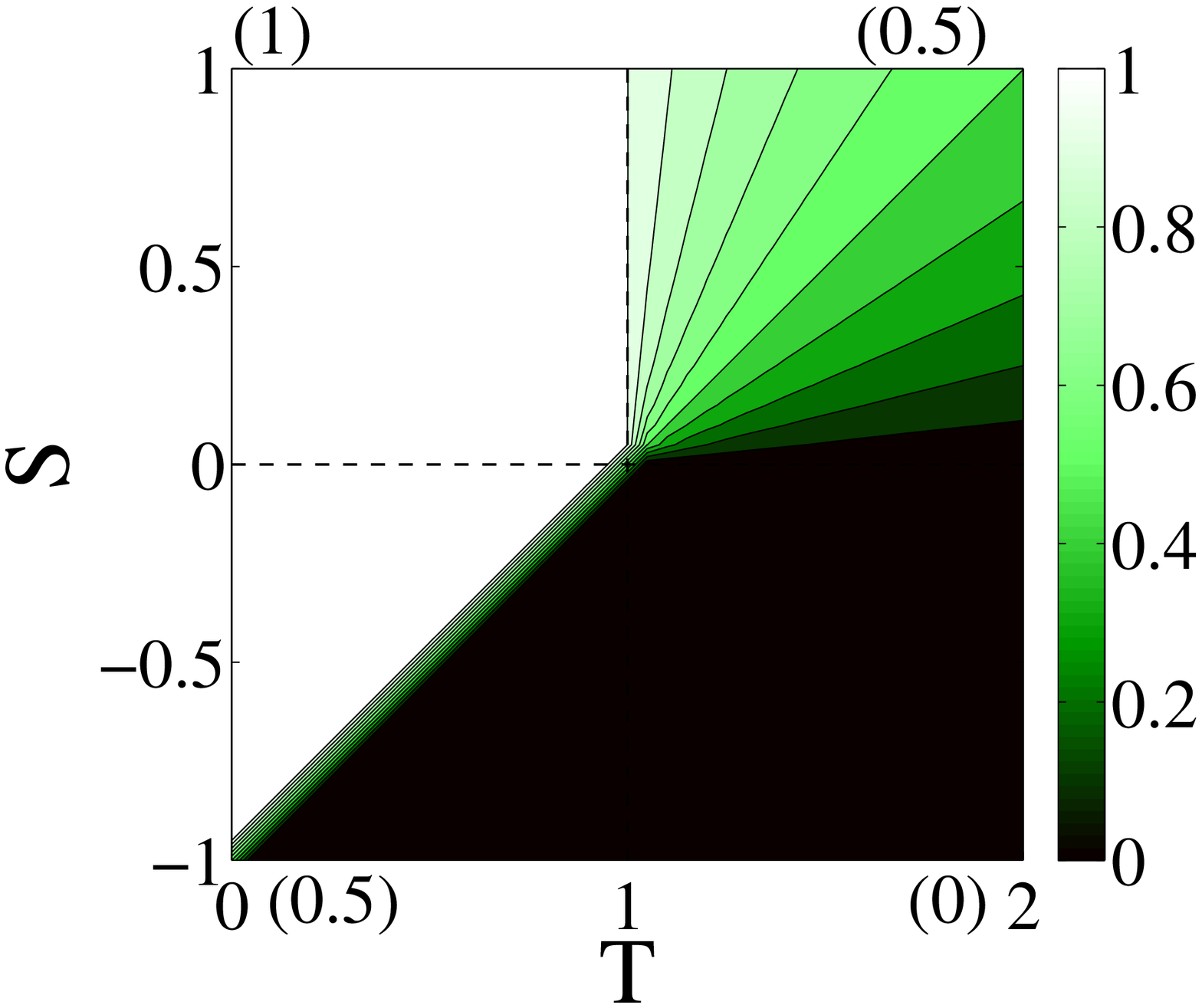} %graphComplet41_rd
\end{tabular}
\caption{(Color online)  Left image: The games phase space (H= Harmony, HD = Hawk-Dove, PD = Prisoner's Dilemma, and SH = Stag Hunt). Right image: Average cooperation over $50$ runs at steady state in a well mixed population (right image). The initial fraction of cooperators is $0.5$ randomly distributed among the graph nodes and the update rule is imitation proportional to the payoff. Lighter tones stand for more cooperation. Figures in parentheses next to each quadrant indicate average cooperation in the corresponding game space. }
 \label{complet}
\end{center}
\end{figure*}

 \begin{figure*}[ht!]
\begin{center}
\begin{tabular} {cccccccc} 
\includegraphics[width=6cm]{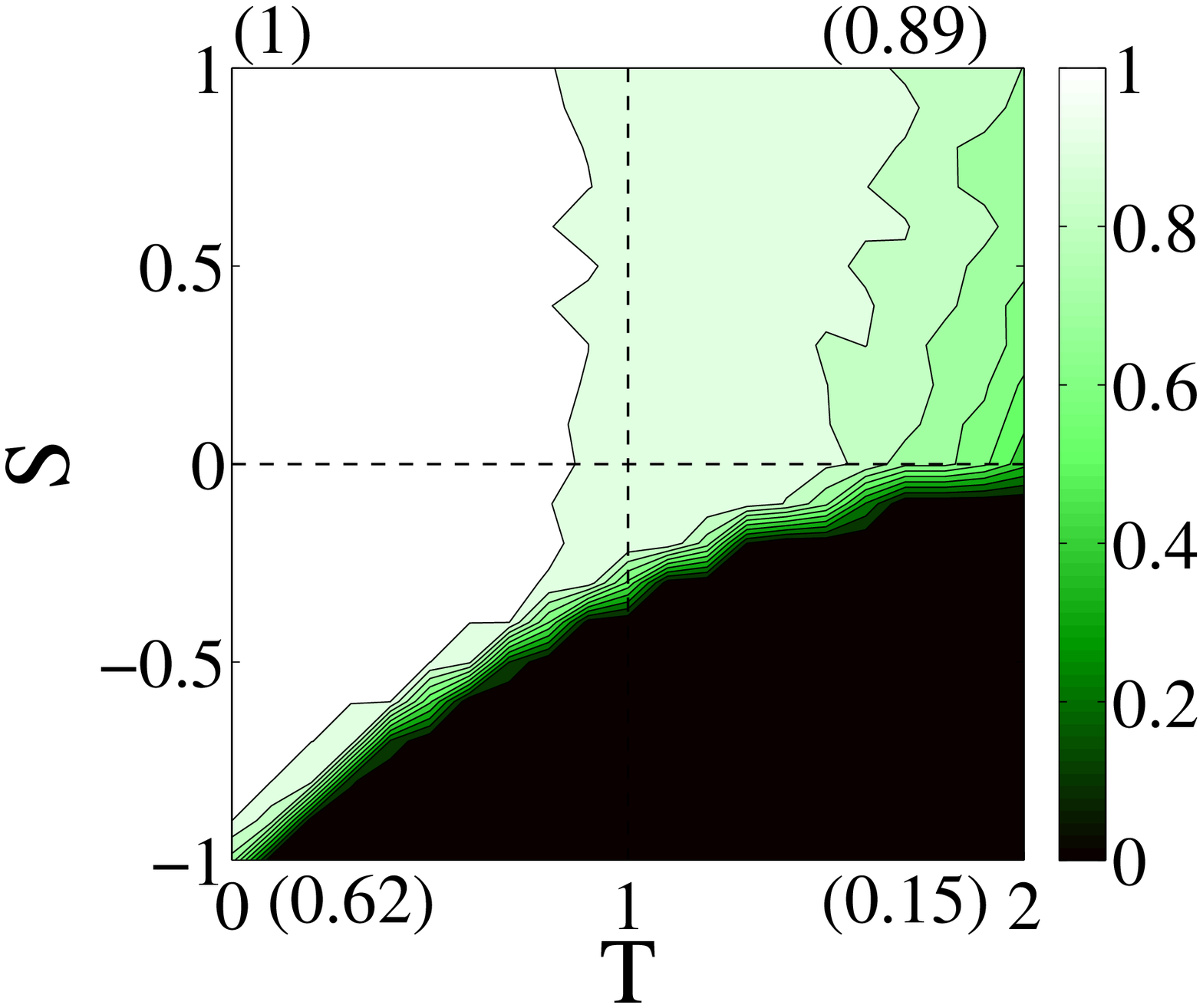} &%BA10000_rd
 \includegraphics[width=6cm]{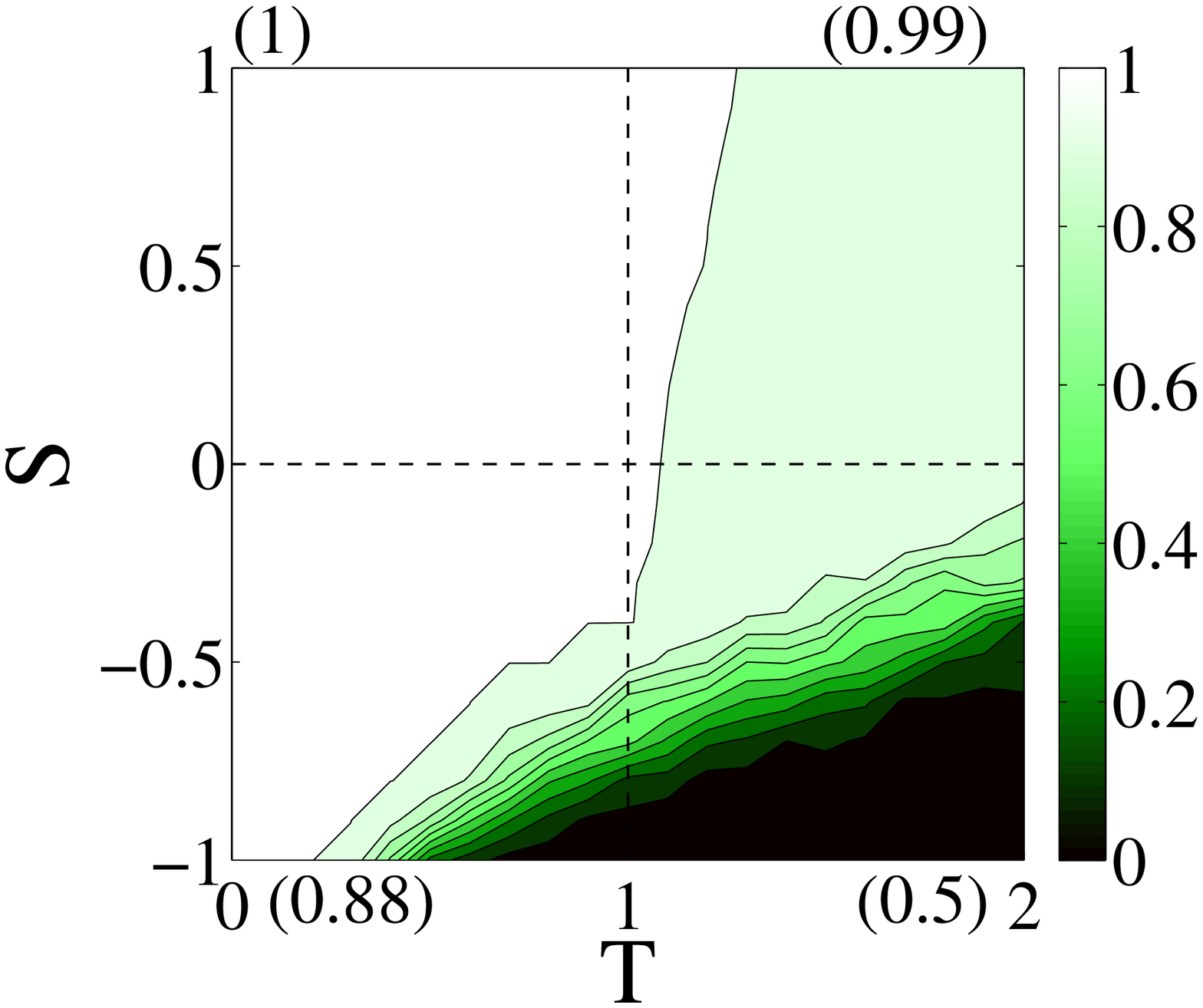} &%BA_ib
 \includegraphics[width=6cm]{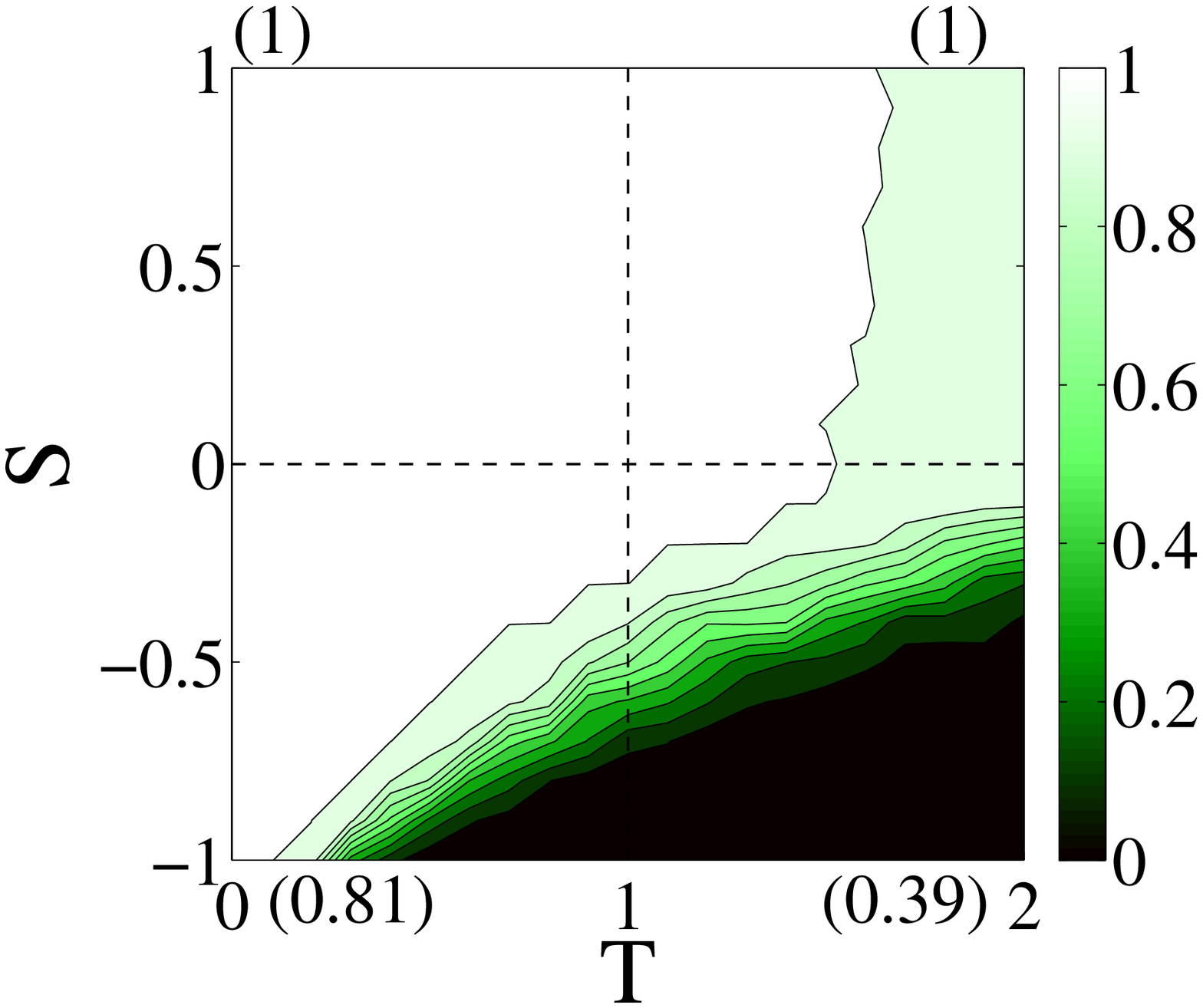} &%BA_ibr
\end{tabular}
%%%%%%%%%%%%%%%%%%%%% !!!!!!!!!!!!!!!!!!!!!!!!!!!!!!!!!!!!!!!!!!!!!!!!!!!!!!!!!!!!!!!!!!!!!!!!!!!!!!!!!!!!!!!!!!!!!!!!!!!!!!!!!!!!!!!!!!!!!!!!!!!!!!!!!!
\caption{(Color online)  Average cooperation over $50$ runs at steady state in  BA networks.  Network size is $N=10000$, $\langle k \rangle  = 8$. The initial fraction of cooperators is $0.5$ randomly distributed among the graph nodes. Lighter tones stand for more cooperation. The update rule is imitation proportional to the payoff (left image), imitation of the best (middle image), randomized imitation of the best (right image). }
 \label{BA}
\end{center}
\end{figure*}

  \begin{figure*}[ht!]
 \begin{center}
\begin{tabular} {cccccccc} 
 \includegraphics[width=6cm]{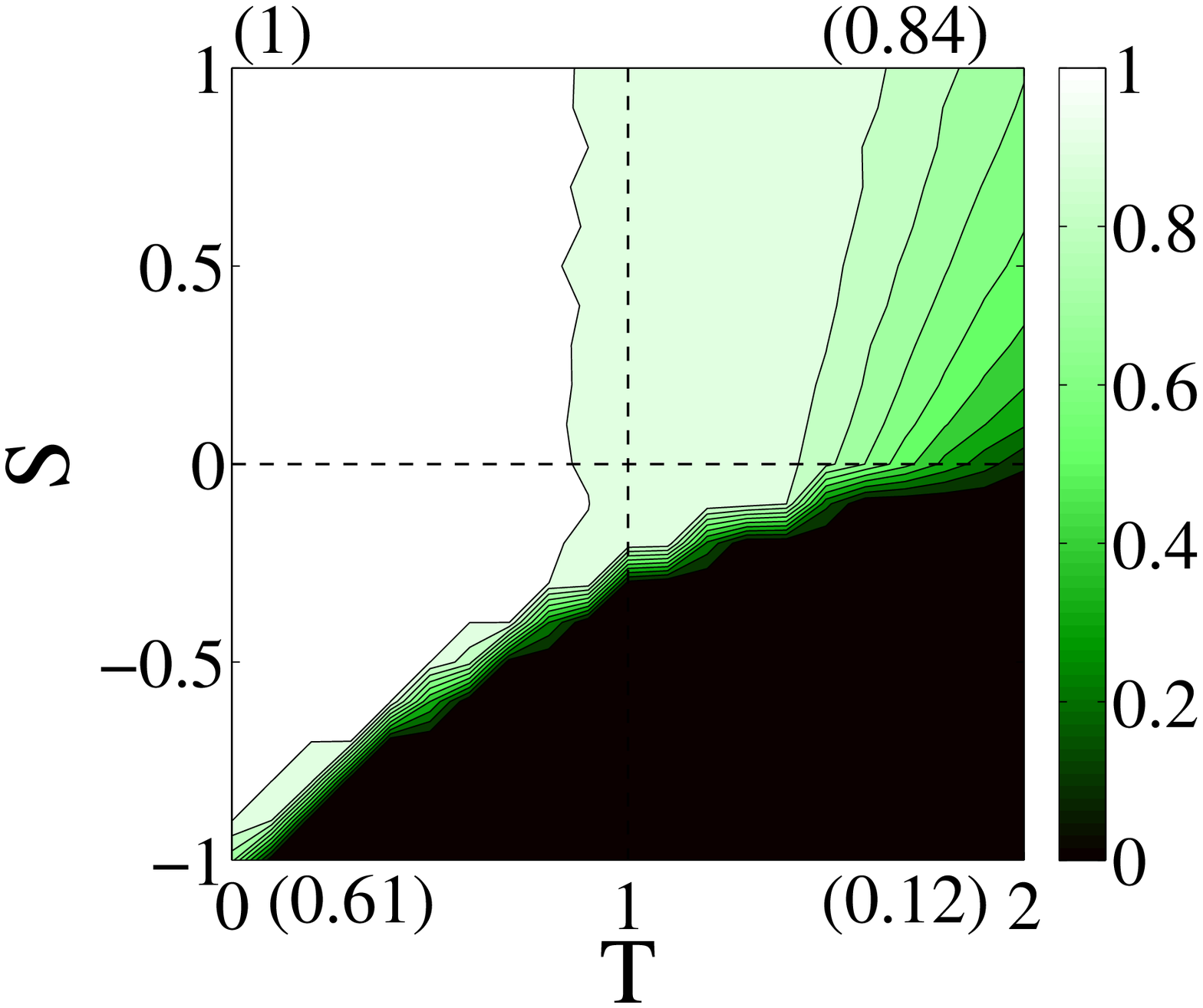} &%ConfigurationModelGamma3_rd
  \includegraphics[width=6cm]{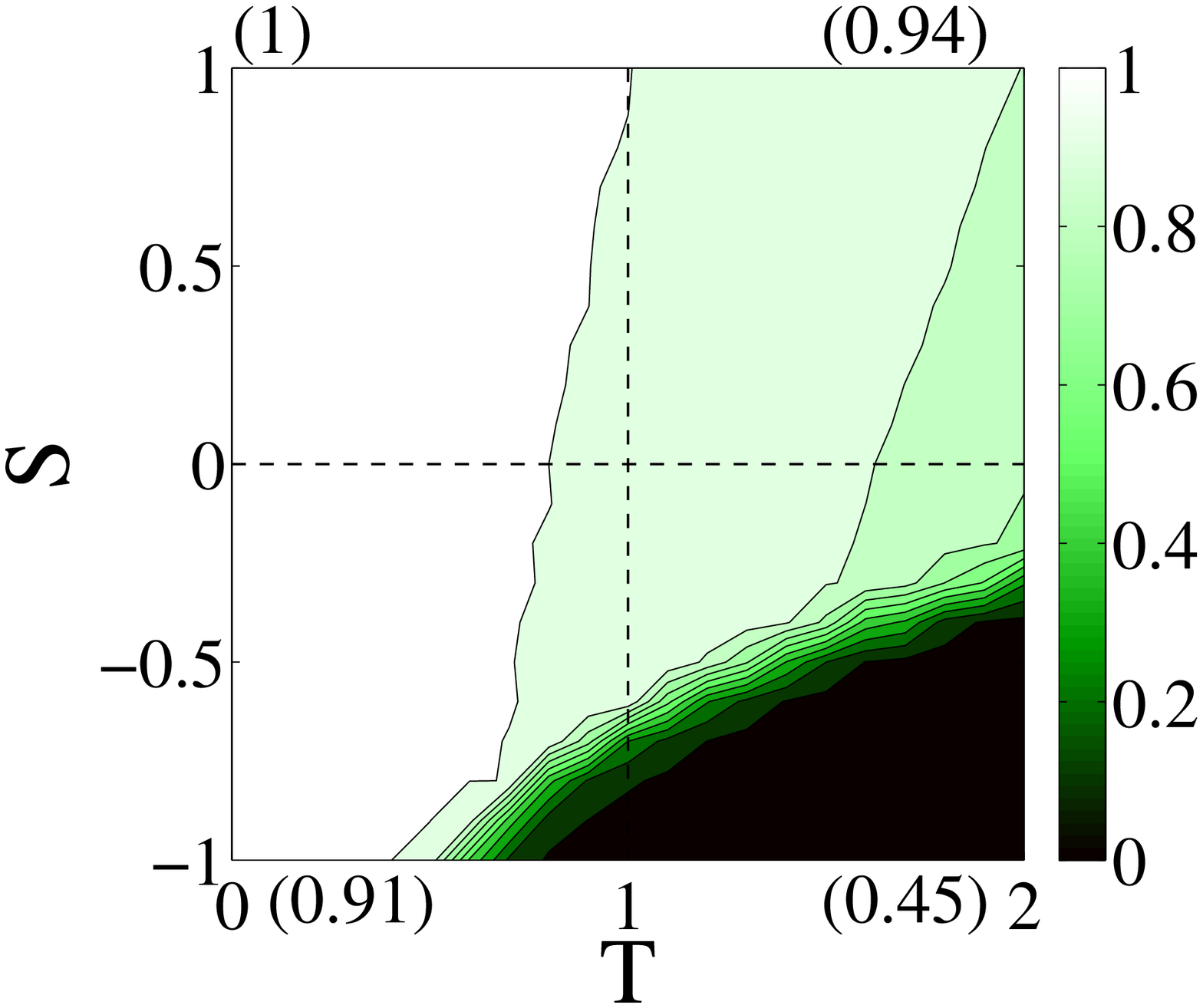} &%ConfigurationModelGamma3_ib
  \includegraphics[width=6cm]{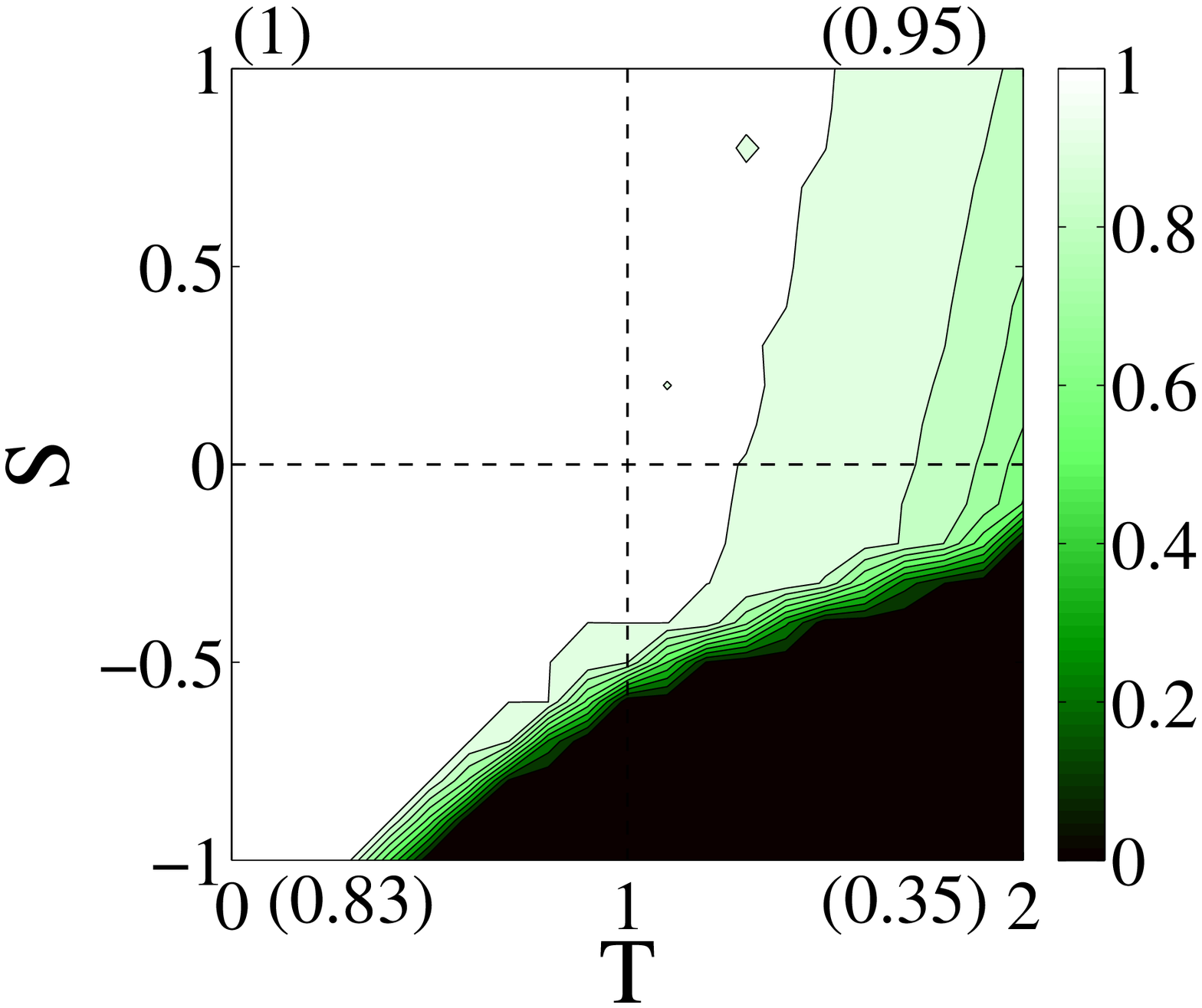} &%ConfigurationModel_ibr

\end{tabular}
\caption{(Color online)  Average cooperation over $50$ runs at steady state on  the configuration model with exponent $\gamma=3.0$, with
$N=10000$,  and $\langle k \rangle  = 8.0$. The initial fraction of cooperators is $0.5$ randomly distributed among the graph nodes. The update rule is imitation proportional to the payoff (left image), imitation of the best (middle image), randomized imitation of the best (right image).}
\label{ConfModel}
\end{center}
 \end{figure*}

  \begin{figure*}[ht!]
\begin{center}
\begin{tabular} {cccccccc} 
\includegraphics[width=6cm]{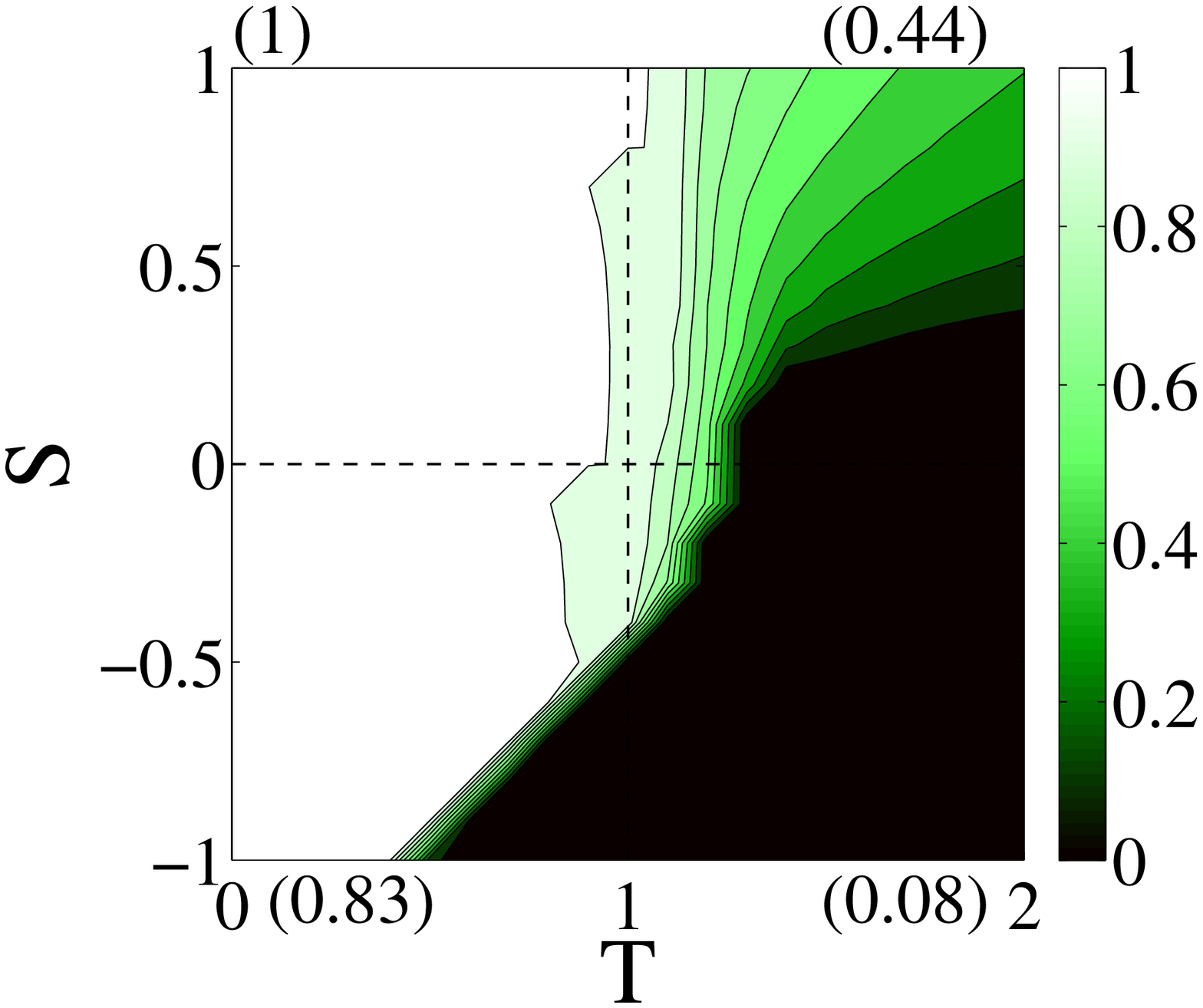} &%RegularLattice10000_rd
\includegraphics[width=6cm]{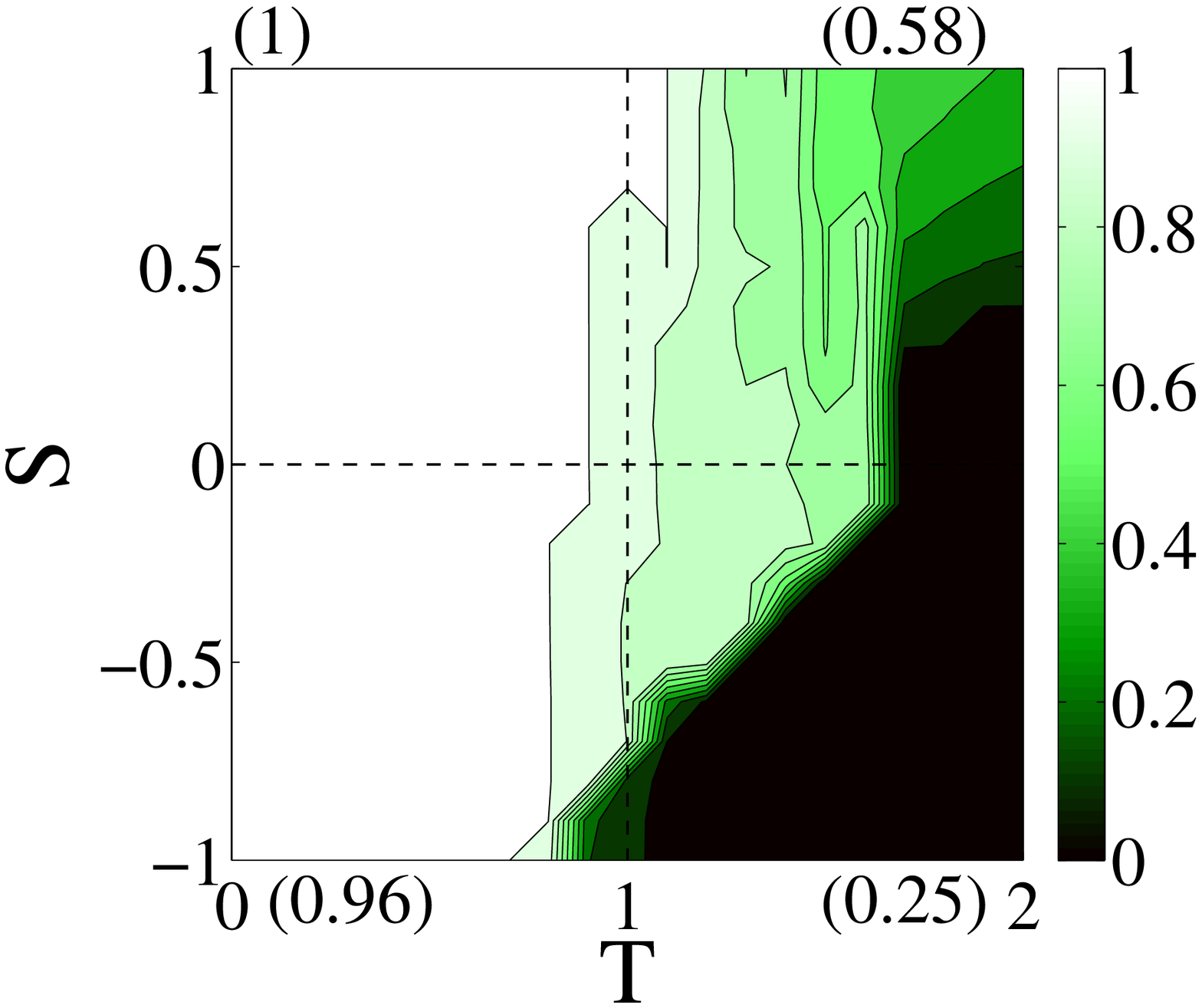} &%RegularLattice_ib
\includegraphics[width=6cm]{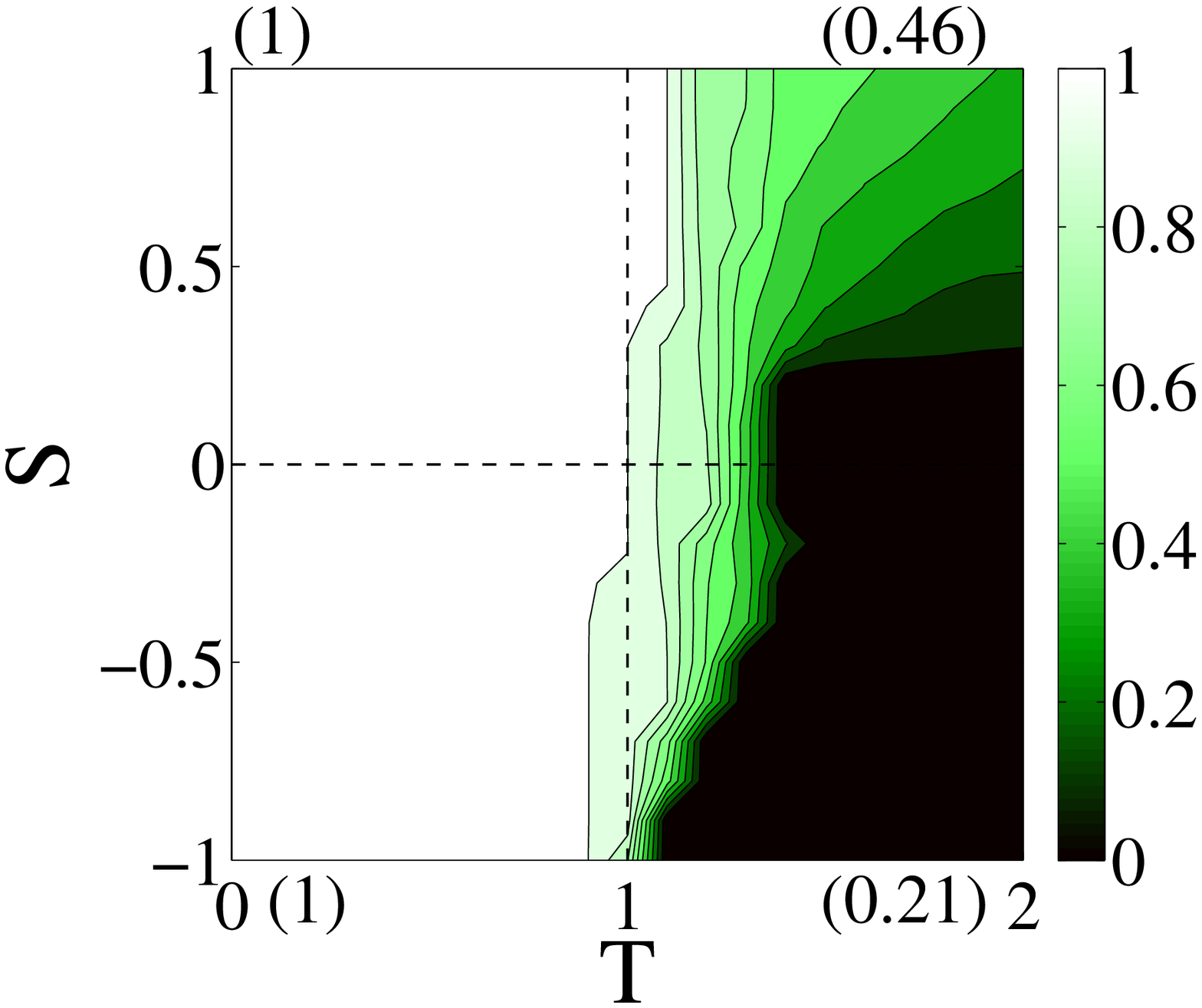} &%RegularLattice_ibr

\end{tabular}
\caption{(Color online)  Average cooperation over $50$ runs at steady state in a regular lattice, the size is $N=10000$, $\langle k \rangle  = 8$. The initial fraction of cooperators is $0.5$ randomly distributed among the graph nodes. Figures in parentheses next to each quadrant indicate average cooperation in the corresponding game space. The update rule is imitation proportional to the payoff (left image), imitation of the best (middle image), randomized imitation of the best (right image). }
 \label{LA}
\end{center}
\end{figure*}

\subsection{Spatial Scale-Free Networks}
\label{SFSN}

The right image of Fig.~\ref{complet} shows cooperation in the well-mixed population case as a baseline to which we refer when evaluating the amount of
cooperation that evolves in network-structured populations.
Scale-free networks, such as the Barab\'asi--Albert (BA)~\cite{alb-baraba-02} and the configuration model (CF)~\cite{ConfModel}, are known to induce high levels of cooperation in HD games and also improve cooperation in the PD and the SH~\cite{santos-pach-06,anxo1}. 
Results on these topologies are shown for future reference in Figs.~\ref{BA} and ~\ref{ConfModel}. 
On the other hand, spatial grids induce high levels of cooperation in SH games but not in PD games, and they reduce the levels of cooperation in the HD games as compared to the well-mixed case~\cite{anxo1}, see Fig.~\ref{LA}.
In order to understand how cooperation is affected by the combination of spatiality and heterogeneous degree distribution, we used a spatial network model with a given degree distribution. 
Our construction is inspired by the one given by
 Rozenfeld et al.~\cite{Rozenfeld2002} and only differs from the latter in the way in which a given node looks for its
 $k$ neighbors.
 We start from a given sequence of target degrees $\{k_1, k_2, \ldots, k_N\}$ and we place the $N$ nodes in
 a regular lattice as in~\cite{Rozenfeld2002}. However,  instead of selecting the nearest neighbors of a given 
 node, the neighbors of a node are chosen in the following way. For each of the $k$ edges of a node we perform a random walk on the underlying lattice starting from this node until we find a free neighbor whose effective degree is less than its target degree and we create a link to that node.  The process is halted when the effective degree of the considered node is equal to the target degree. Since it is possible that a node has already cumulated edges up to its target degree, we fix a maximum $m = N$ for the number of random walk steps for the construction of one edge. We used a scale-free distribution $p(k) \propto k^{-\gamma}$ with exponent $\gamma \in \{2.0,\:3.0,\:4.0\}$. In order to keep a constant mean degree  $\langle k \rangle = 8$, the lower bound of the scale-free degree sequence was shifted. Thus, as $\gamma$ increases the distribution becomes more peaked around $\langle k \rangle$. The
 network model just described, called SFSN,  gives very similar results on cooperation as the one of Rozenfeld et al.; thus we show only those concerning our model.
 
  \begin{figure*}[ht!]
 \begin{center}
\begin{tabular} {cccccccc} 
\includegraphics[width=4.3cm]{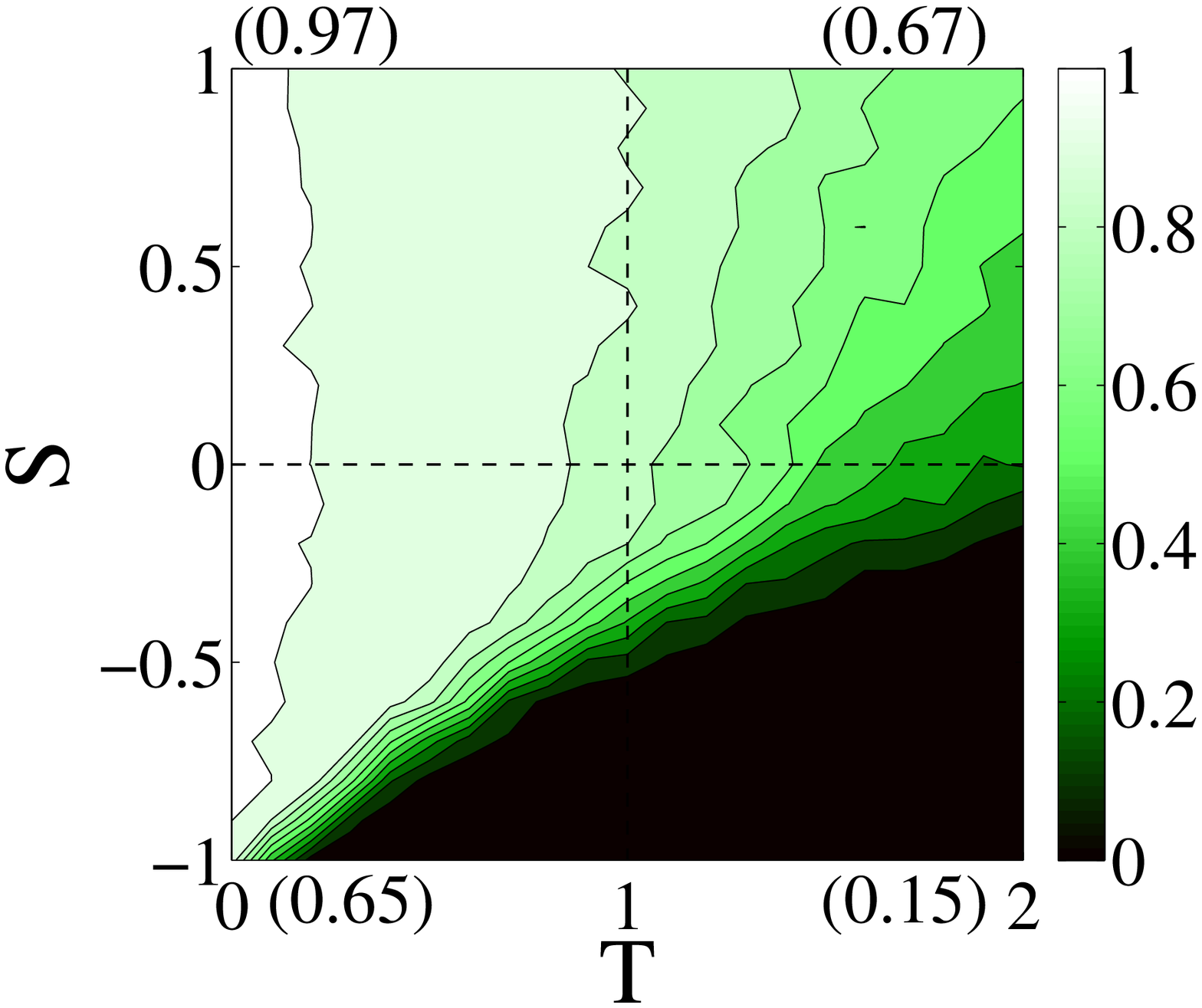} &%_spatialScaleFreeGamma2T40000_rd
 \includegraphics[width=4.3cm]{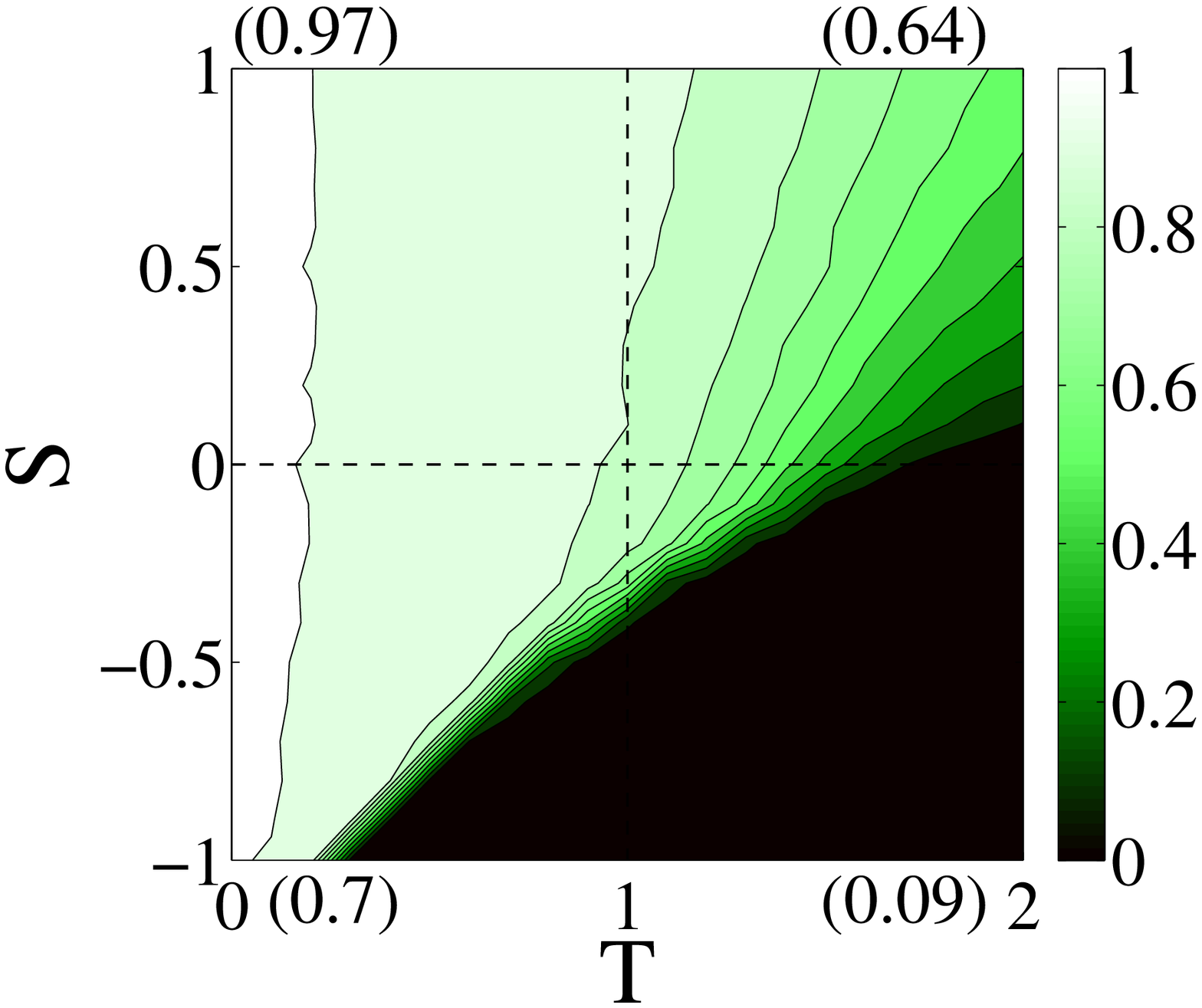} &%_spatialScaleFreeGamma3T40000_rd
 \includegraphics[width=4.3cm]{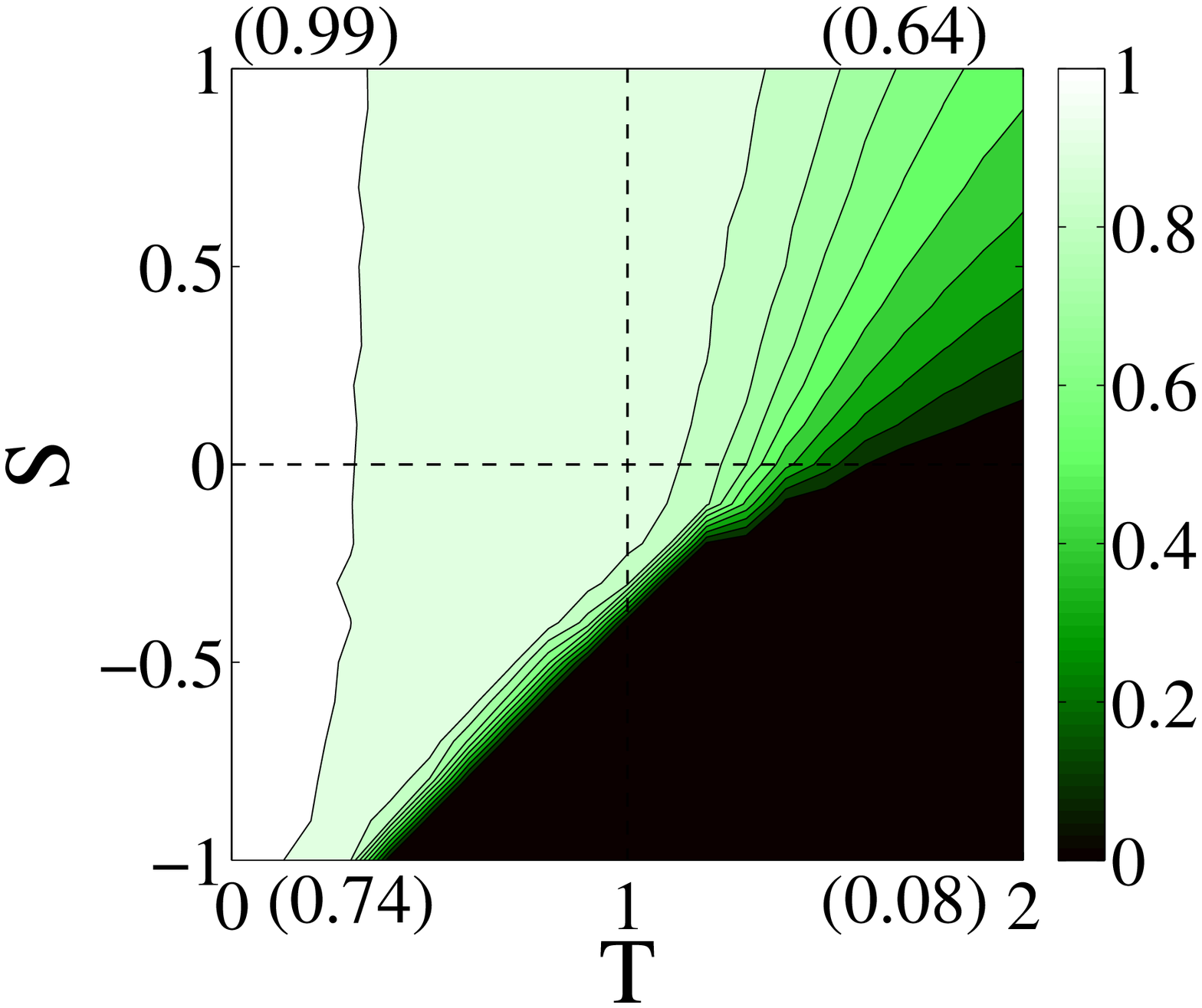} &%_spatialScaleFreeGamma4T40000_rd
 \includegraphics[width=4.3cm]{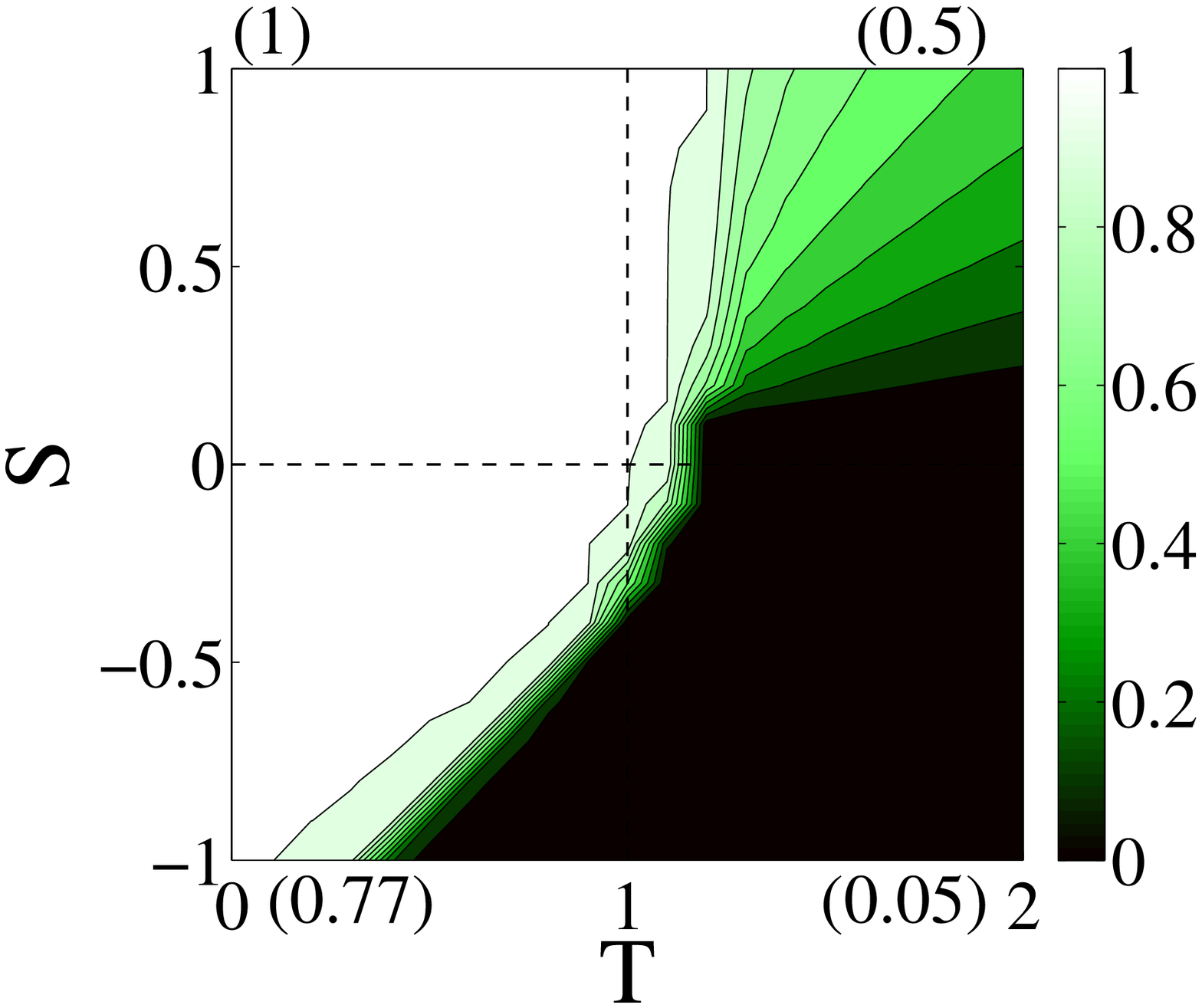}%geomGraphK20N10000_rd 
\end{tabular}
\caption{(Color online)  Average cooperation over $50$ runs at steady state on  SFSN networks (see text). Size is $N=10000$, $\langle k \rangle  = 8$ and $\gamma = 2.0$ (left image), $3.0$ (middle image), and $4.0$ (right image). The rightmost image corresponds to
random geometric graphs with $\langle k \rangle  = 20$. The initial fraction of cooperators is $0.5$ randomly distributed among the graph nodes and the update rule is imitation proportional to the payoff.}
\label{SpatialScaleFree}
\end{center}
 \end{figure*}

\begin{figure*}[ht!]
 \begin{center}
\begin{tabular} {cccccccc} 
\includegraphics[width=4.3cm]{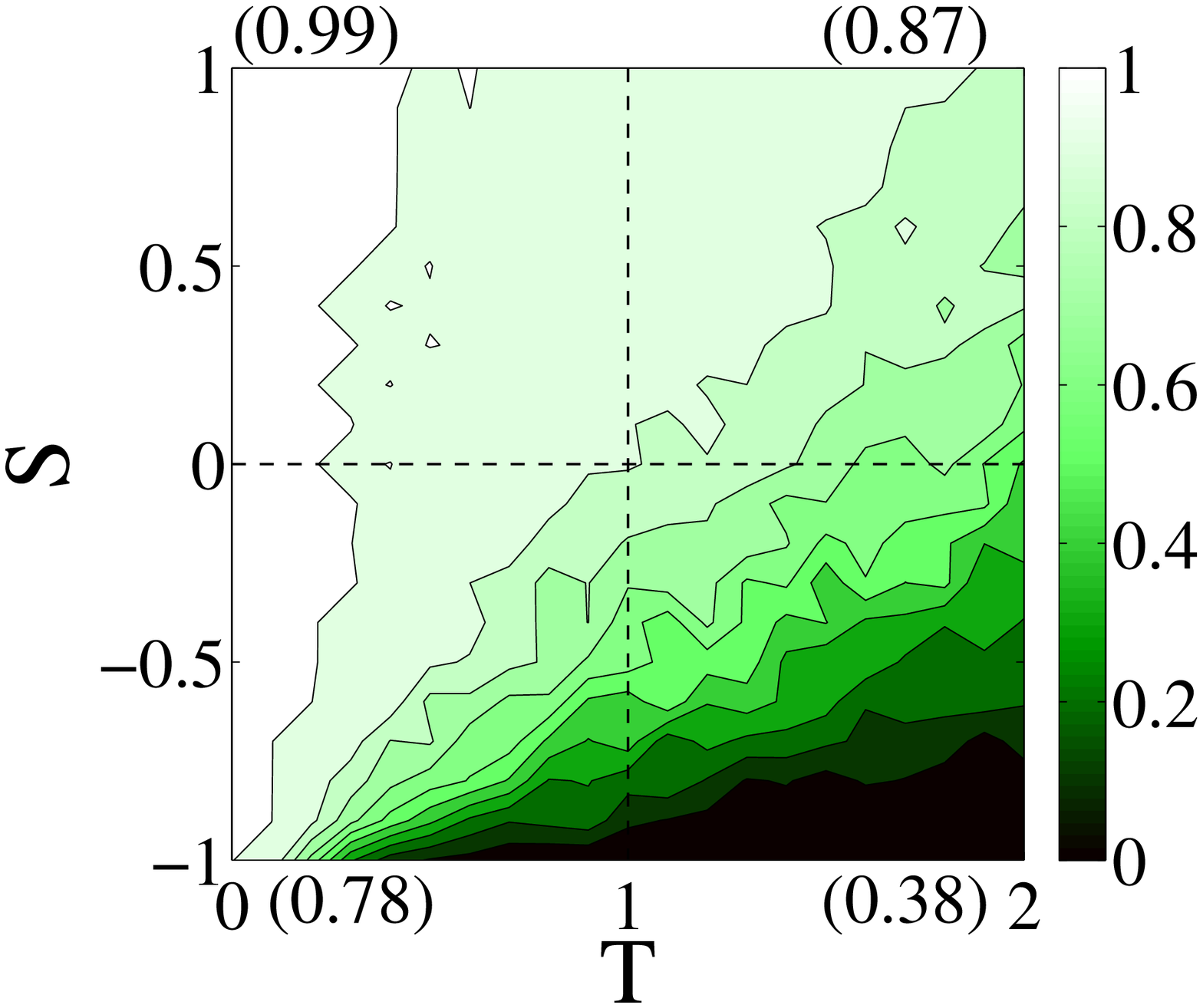} &%_spatialScaleFreeGamma2_ib
 \includegraphics[width=4.3cm]{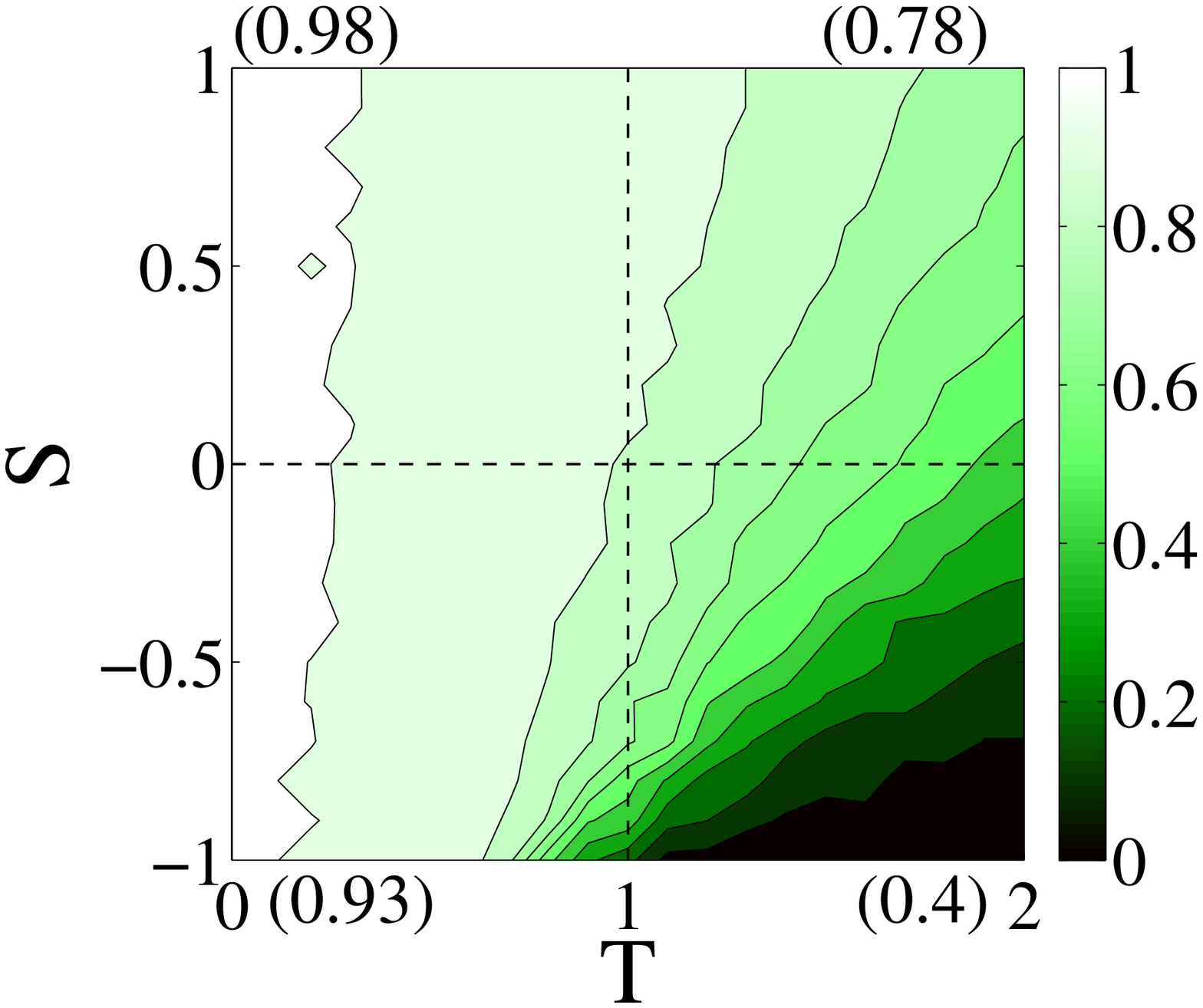} &%_spatialScaleFreeGamma3_ib
 \includegraphics[width=4.3cm]{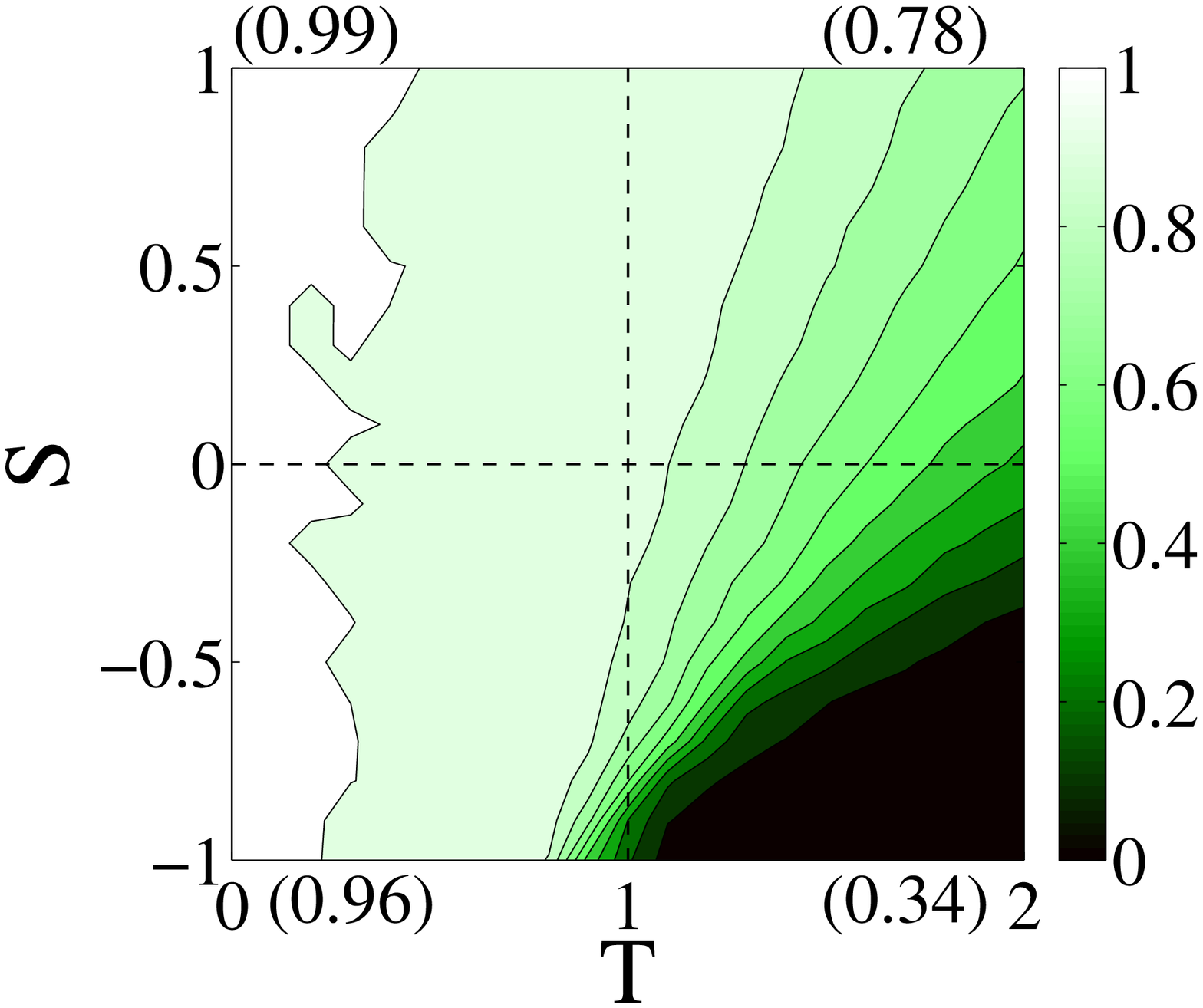} &%_spatialScaleFreeGamma4_ib
  \includegraphics[width=4.3cm]{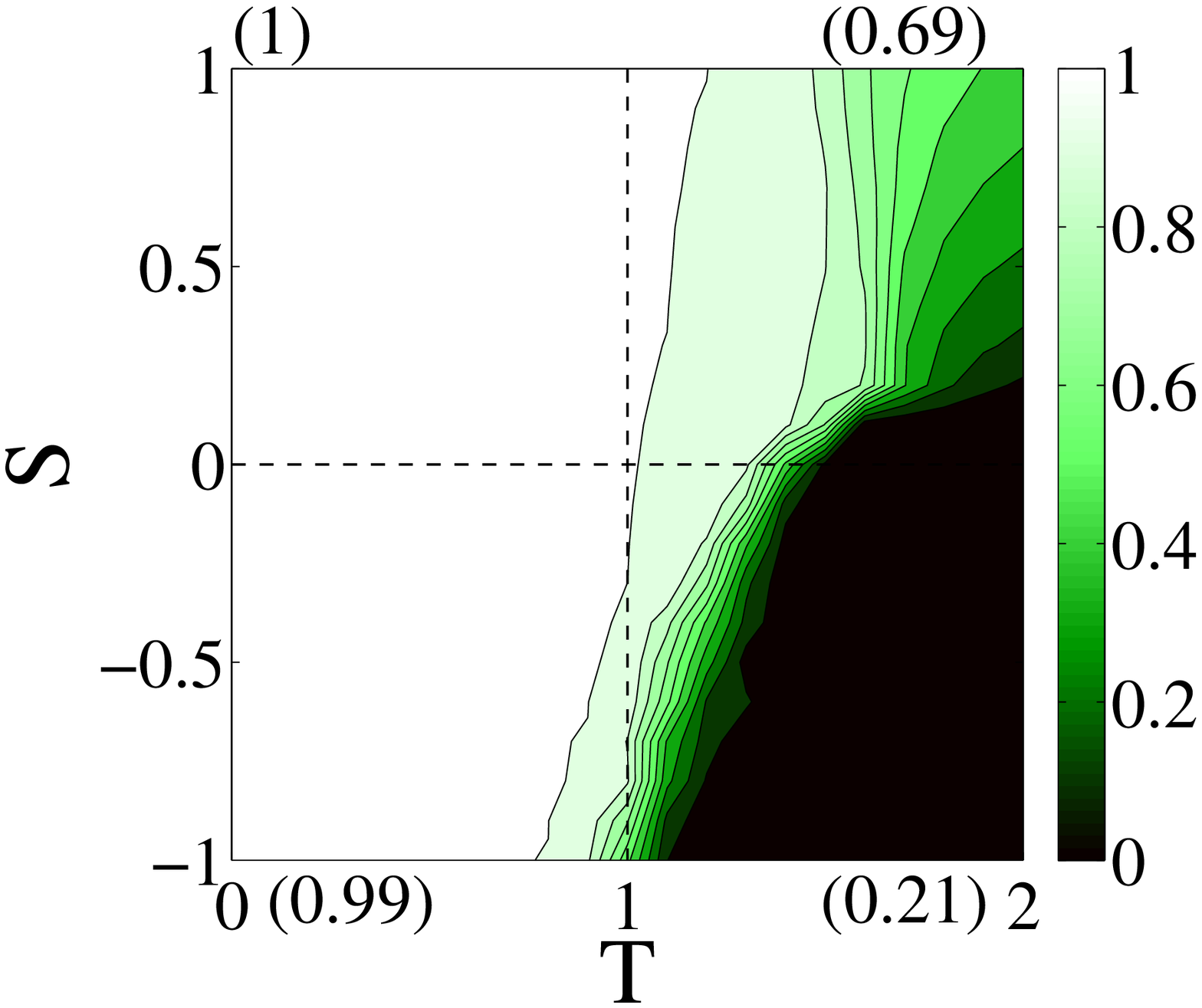} &%RandGeomk20_ib
\end{tabular}
\end{center}
\caption{(Color online)  Average cooperation over $50$ runs at steady state on  SFSN networks with imitation of the best update rule. Size is $N=10000$, $\langle k \rangle  = 8$ and $\gamma = 2.0$ (left image), $3.0$ (middle image), $4.0$ (right image). The rightmost image corresponds to
random geometric graphs with $\langle k \rangle  = 20$. The initial fraction of cooperators is $0.5$ randomly distributed among the graph nodes.}
 \label{SpatialScaleFree_ib}
\end{figure*}

For comparison purposes, and by analogy with  Erd\"os-R\'enyi random graphs~\cite{newman-book} in relational
networks, we take as a spatial baseline case the random geometric graph (RGG).
Random geometric graphs are constructed as follows~\cite{Dall,Barthelemy}. $N$ nodes are placed randomly on a subset of $\mathbb{R}^n$; then two nodes are linked if their distance is less than a constant $r$. The resulting graph has a binomial degree distribution which tends to a Poisson degree distribution as  $N\to \infty$ and $r\to 0$~\cite{Dall,Barthelemy},
 with $\langle k \rangle=Nq(r)$ and $q(r)=\pi r^2/S$, for $n=2$, is the probability that a node is in a disk of radius $r$. $S$ is the total surface.

Average cooperation on SFSNs  with RD and IB update rules 
for $\gamma=2.0, 3.0, 4.0$  are shown in Figs.~\ref{SpatialScaleFree} and~\ref{SpatialScaleFree_ib}. The first remark is that spatial scale-free networks are slightly less conducive to cooperation than the corresponding BA and CF relational networks. This can be seen by 
comparing the first images of Fig.~\ref{BA} and Fig.~\ref{ConfModel}, which corresponds to the BA and CF cases, with the second image of Fig.~\ref{SpatialScaleFree}
which refers to SFNSs with $\gamma=3$. Nevertheless, it can be seen that SFNSs do favor cooperation especially in
the HD and the PD space with respect to the RGG case depicted in Fig.~\ref{SpatialScaleFree}, rightmost
image.
It can also be observed that the gains in the transition between cooperation and defection that is
apparent  in the SH games with increasing $\gamma$, are partially 
offset for low $\gamma$. 
The comparison with the RGG case (rightmost image) shows that cooperation levels
tend to those of the random graph case with increasing $\gamma$, except for the HD quadrant where
the RGG topology causes some cooperation loss.
As in non-spatial networks~\cite{anxo1}, the imitation of the best neighbor strategy update rule
is more noisy and gives rise, in general, to somewhat higher levels of cooperation. Results with IBR rule are similar to those with IB and are not presented.

%-------------------------------------------------------------------------------------------------------------------------------------------------------------------
\subsection{Apollonian Networks: A Spatial Scale-Free Model with Higher Cooperation}
\label{AN}

\begin{figure*}[ht!]
 \begin{center}
\begin{tabular} {cccccccc} 
\includegraphics[width=5cm]{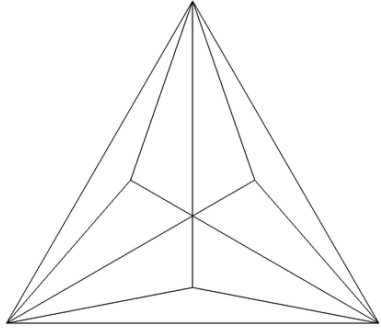} &%ApolloniusSimple
\end{tabular}
\caption{(Color online)Apollonian network after two generations (see text).
}
\label{ApolloniusDrawing}
\end{center}
 \end{figure*}

An interesting case of  scale-free spatial networks are the Apollonian networks (AN)~\cite{Apollonius2005Herrmann} for which
we show in this section that they lead to high levels of cooperation.
Apollonian networks are constructed by linking adjacent circles in Apollonian packings. In the simplest case, an Apollonian packing is built by starting from three tangent circles, adding a smaller circle tangent to the three previous ones, and iterating the process for each new hole between three circles (see Fig.~\ref{ApolloniusDrawing} and~\cite{Apollonius2005Herrmann}). Our sample Apollonian networks have been obtained in this way after nine iterations and are of size $N=9844$ nodes.
AN belong to a class of networks that are scale-free, small-world, planar and embedded in Euclidean space. The degree-distribution exponent is $\gamma = 2.585$ and the clustering coefficient $C_{cl} = 0.83$~\cite{Apollonius2005Herrmann}.

\begin{figure*}[ht!]
 \begin{center}
\begin{tabular} {cccccccc} 
 \includegraphics[width=4.3cm]{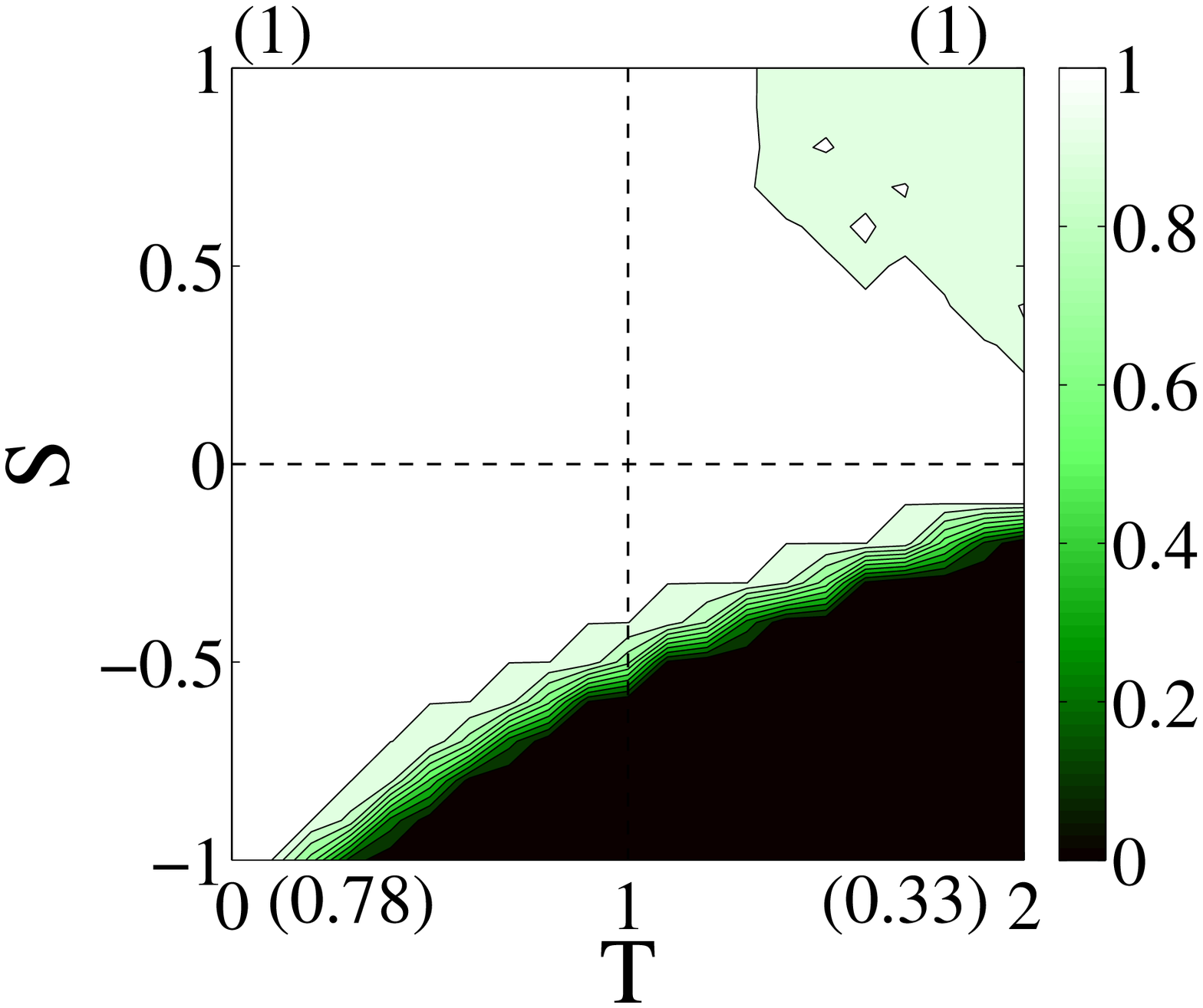} &%Apollonianit9T20000_rd
 \includegraphics[width=4.3cm]{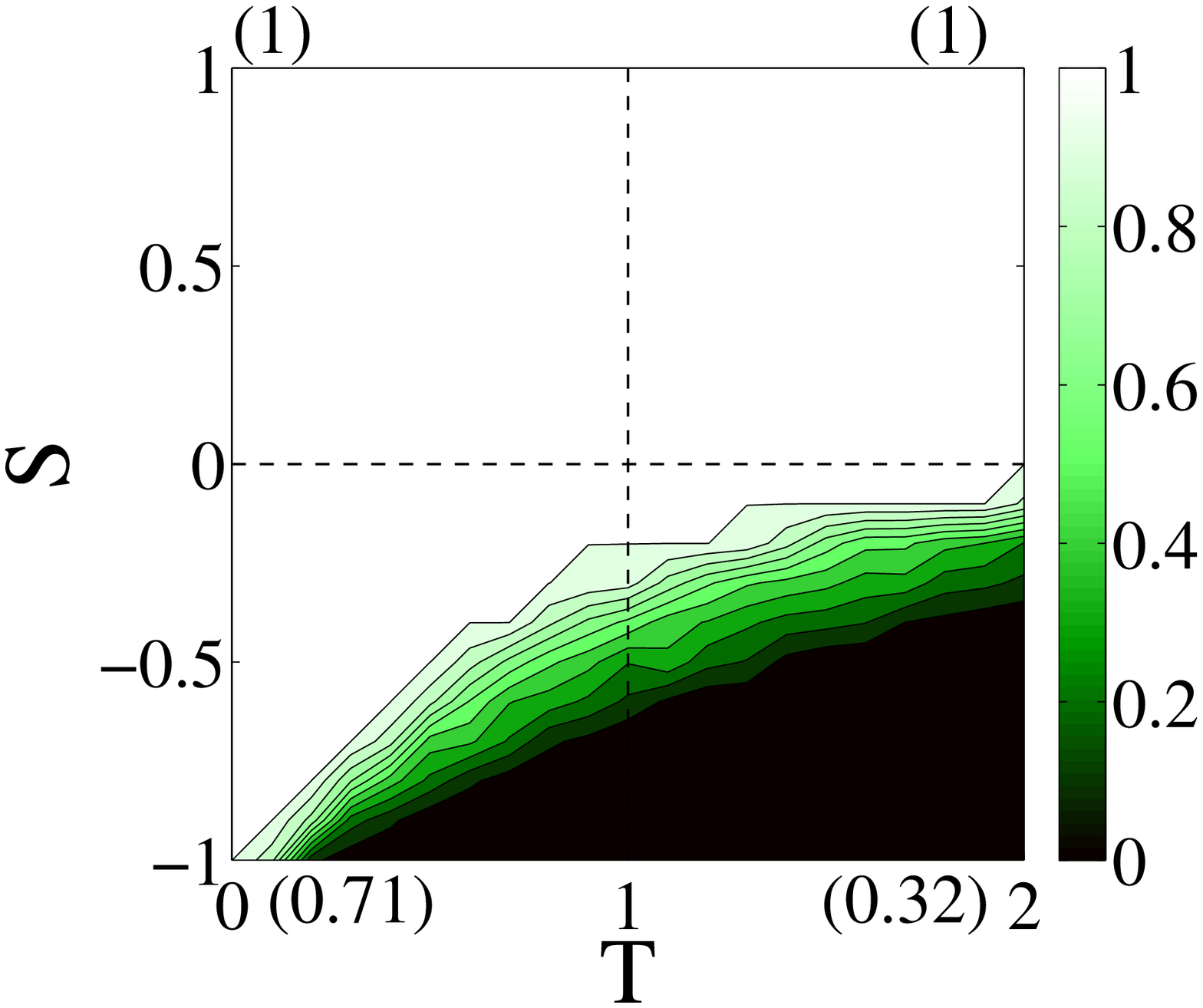} &%ApollonianBeta1_fermi
 \includegraphics[width=4.3cm]{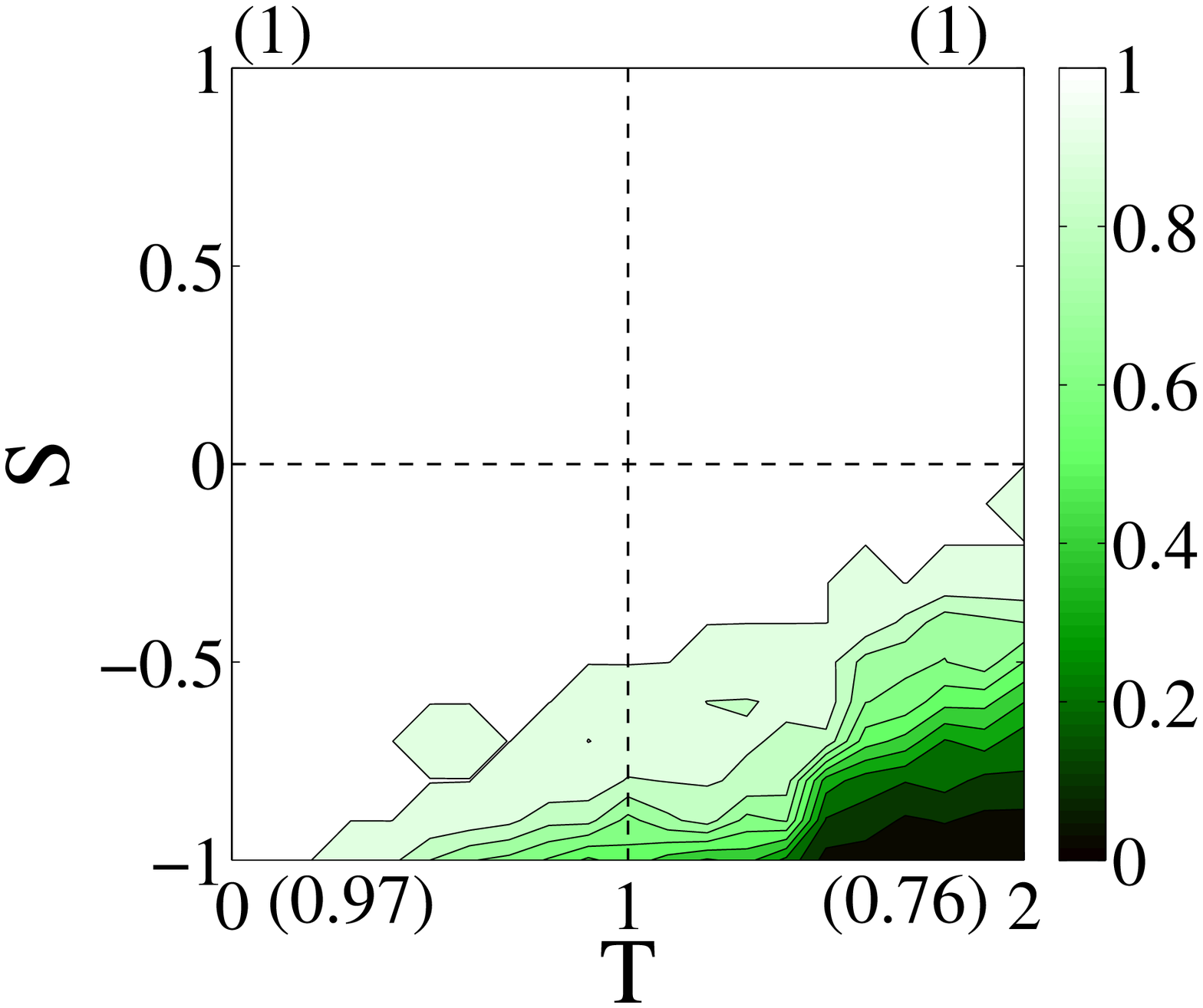} &%apollonianIt9_ib
  \includegraphics[width=4.3cm]{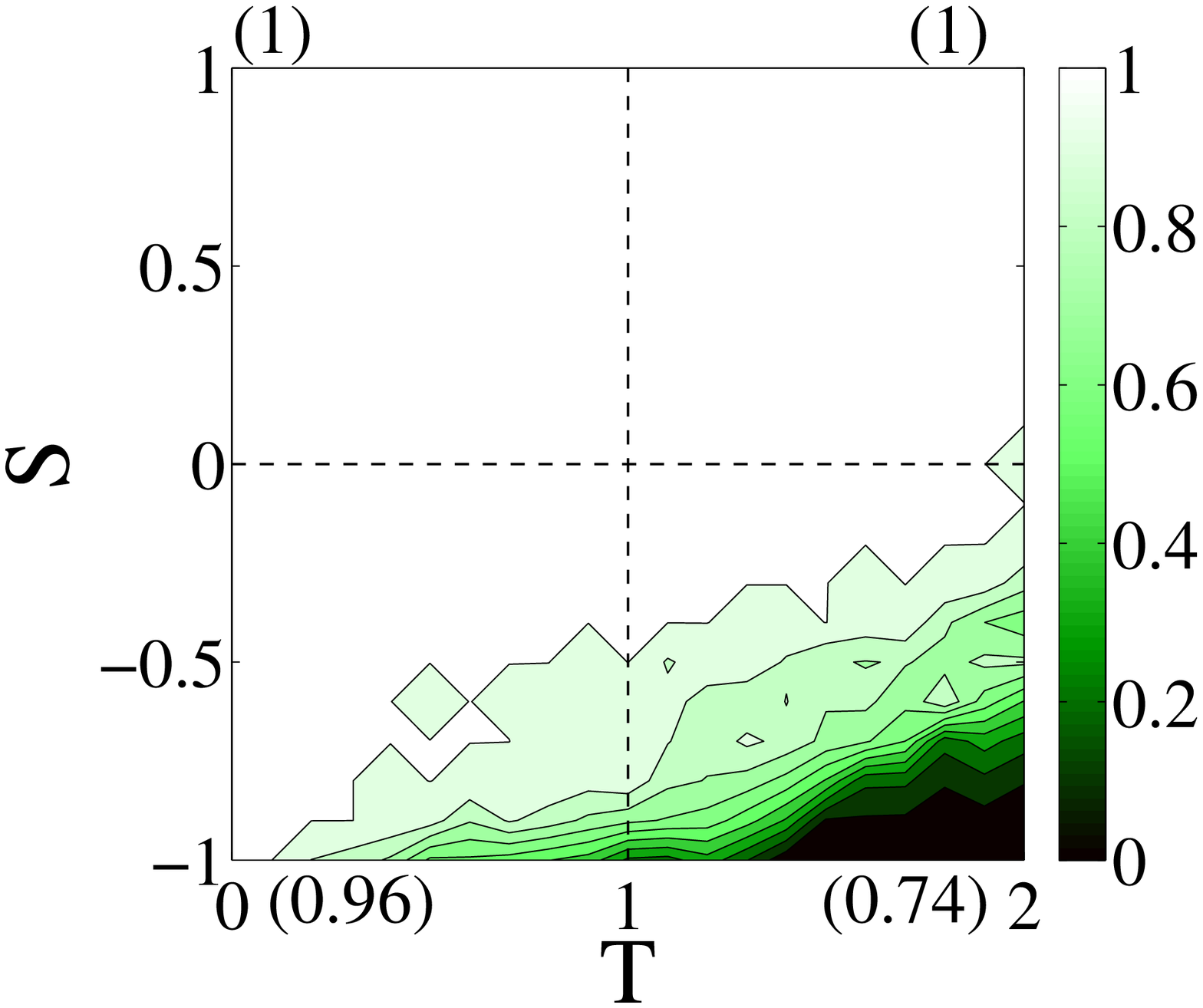} &%apollonianIt9T20000_ibr
\end{tabular}
\caption{(Color online)  Average cooperation over $50$ runs at steady state on the Apollonian network of size
 $N=9844$, and $\langle k \rangle  \simeq 6.0$. The initial fraction of cooperators is $0.5$ randomly distributed among the graph nodes. 
 From left to right the strategy update rules are: replicator dynamics, Fermi rule, imitation of the best, and probabilistic imitation of the
 best.}
\label{ApolloniusSimple}
\end{center}
 \end{figure*}

In a recent work, Yang et al.~\cite{ApolloniusCoop2011} have shown that
Apollonian networks foster cooperation on the weak prisoner's dilemma ($R = 1, P=0, S=0, T\in[0,3]$) using update proportional
to payoff. The space covered is thus just the segment at the frontier between HD and PD. Here,
 with the aim of extending the scope of the study, we sample the full ST-plane.
Our results are summarized in Fig.~\ref{ApolloniusSimple}. They show that the AN topology is more conducive to cooperation than SFNS and
BA networks in the HD games, but also in the SH games, by shifting  to the right the transition from cooperation to defection at $S=-1$,
as in other spatial networks. 
However, the amount of cooperation gain depends on the strategy update rule. Replicator dynamics and the Fermi rule
(first and second image from the left respectively) have
a similar behavior, and are also analytically close (if the exponent $\beta$ of the Fermi rule is between $1$ and $10$). On the other hand,
the rule that prescribes straight  imitation of the best (third image), and the rule that imitates the best neighbor probabilistically
(rightmost image) perform better. 
Intuitively, the first two rules choose a very good neighbor to imitate less often than the latter two, especially when compared with
deterministic imitation of the best. This could favor the latter rules in an Apollonian network when some cooperators surrounded by
a majority of cooperators have gained a foothold on several hubs.

In ~\cite{ApolloniusCoop2011}, the authors discuss the topology features that induce the high cooperation levels. They point out, beside other facts, the presence of connections between hubs and that there exist nodes with high $g_i$ and $U_i$, $g_i$ being the degree gradient between a node $i$ and its neighbors $\{V_i\}$, $g_i=\sqrt{\frac{1}{k_i}\sum_{j\in V_i}(k_i-k_j)^{2}}$ and $U_i=k_i*g_i$.  By transforming the network they show that these features are linked to high levels of cooperation. They point out the high clustering coefficient and explain that clustering increases cooperation on the reduced PD games as shown in~\cite{cluster2008}.

%\enlargethispage{\baselineskip}

\subsection{High Levels of Cooperation on Lattice and Derived Structures}
\label{supergrid}
\begin{figure*}[ht!]
\begin{center}
\includegraphics[width=5.5cm]{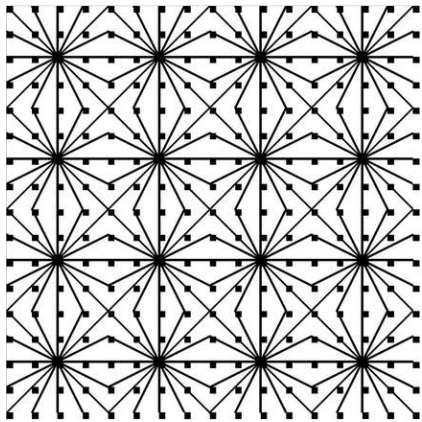}%supergridWhite 
\caption{(Color online)  Lattice topology with two kind of nodes. Each hub is fully connected to a square neighborhood of side $5$ and to the $4$ nearest hubs.  }
 \label{supergridimage}
\end{center}
\end{figure*}

\begin{figure*}[ht!]
\begin{center}
\begin{tabular} {cccccccc} 
 \includegraphics[width=4.3cm]{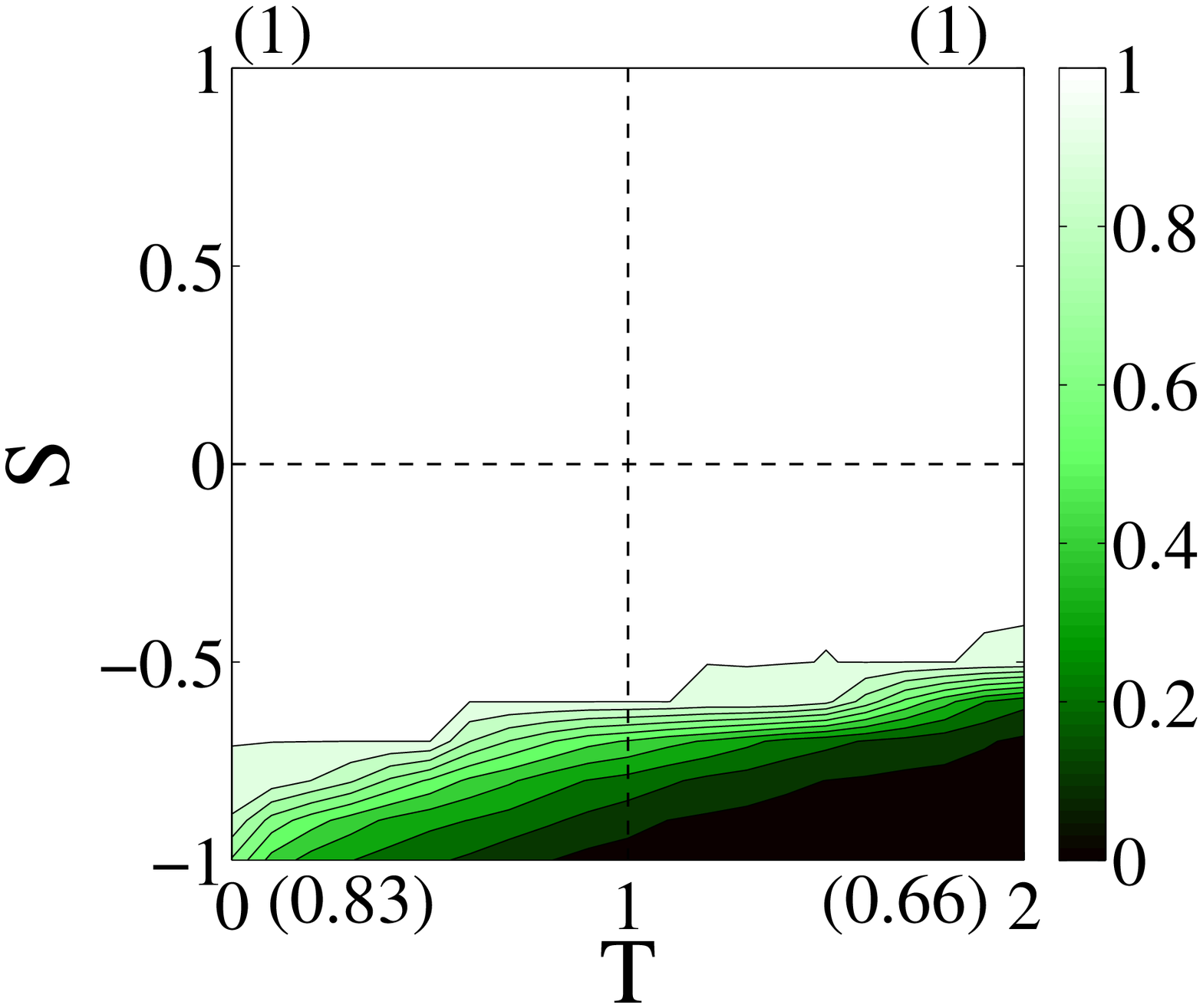} &%supergrid_rd
 \includegraphics[width=4.3cm]{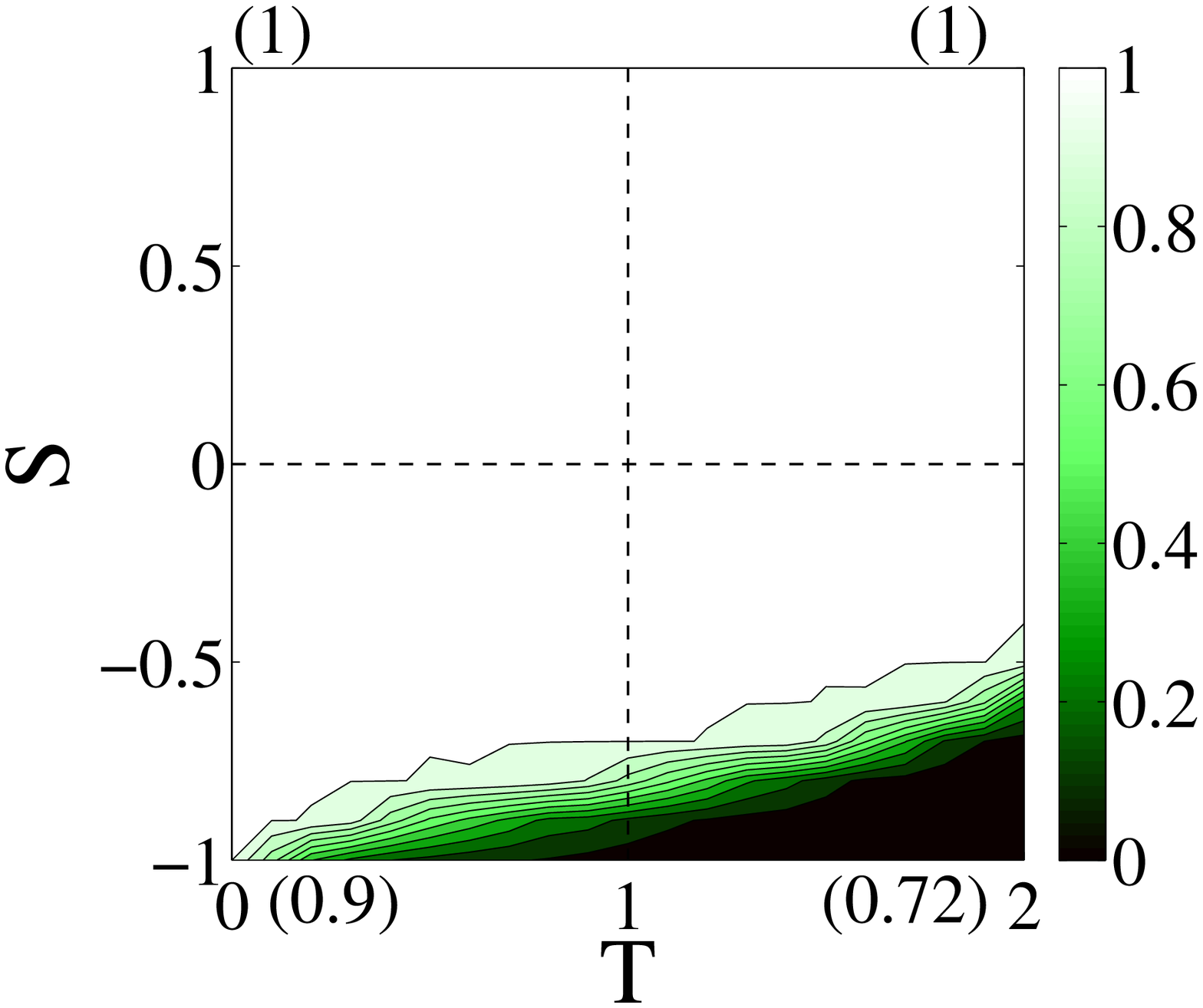} &%supergridBeta1_fermi
\includegraphics[width=4.3cm]{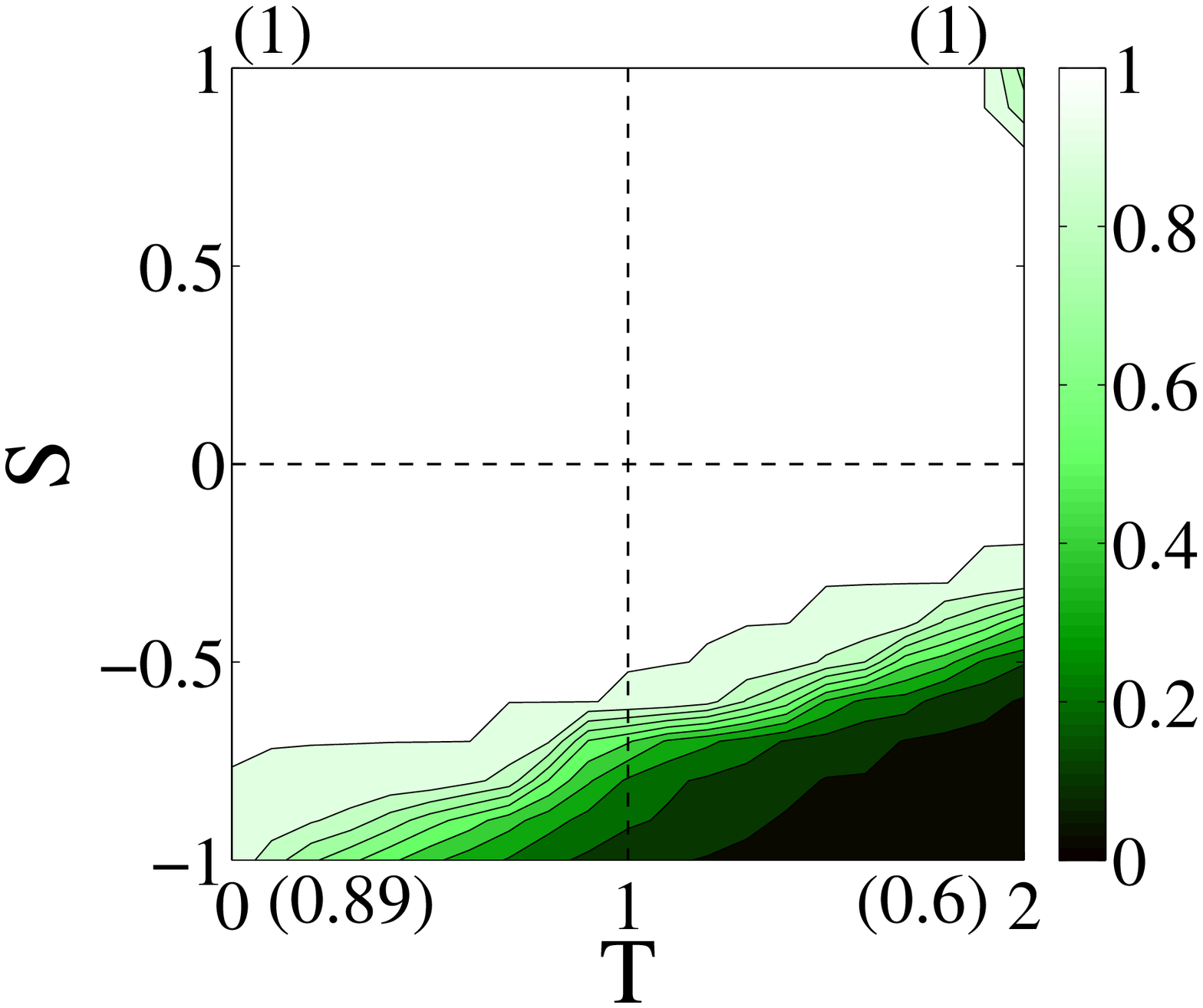} &%supergrid_ib
\includegraphics[width=4.3cm]{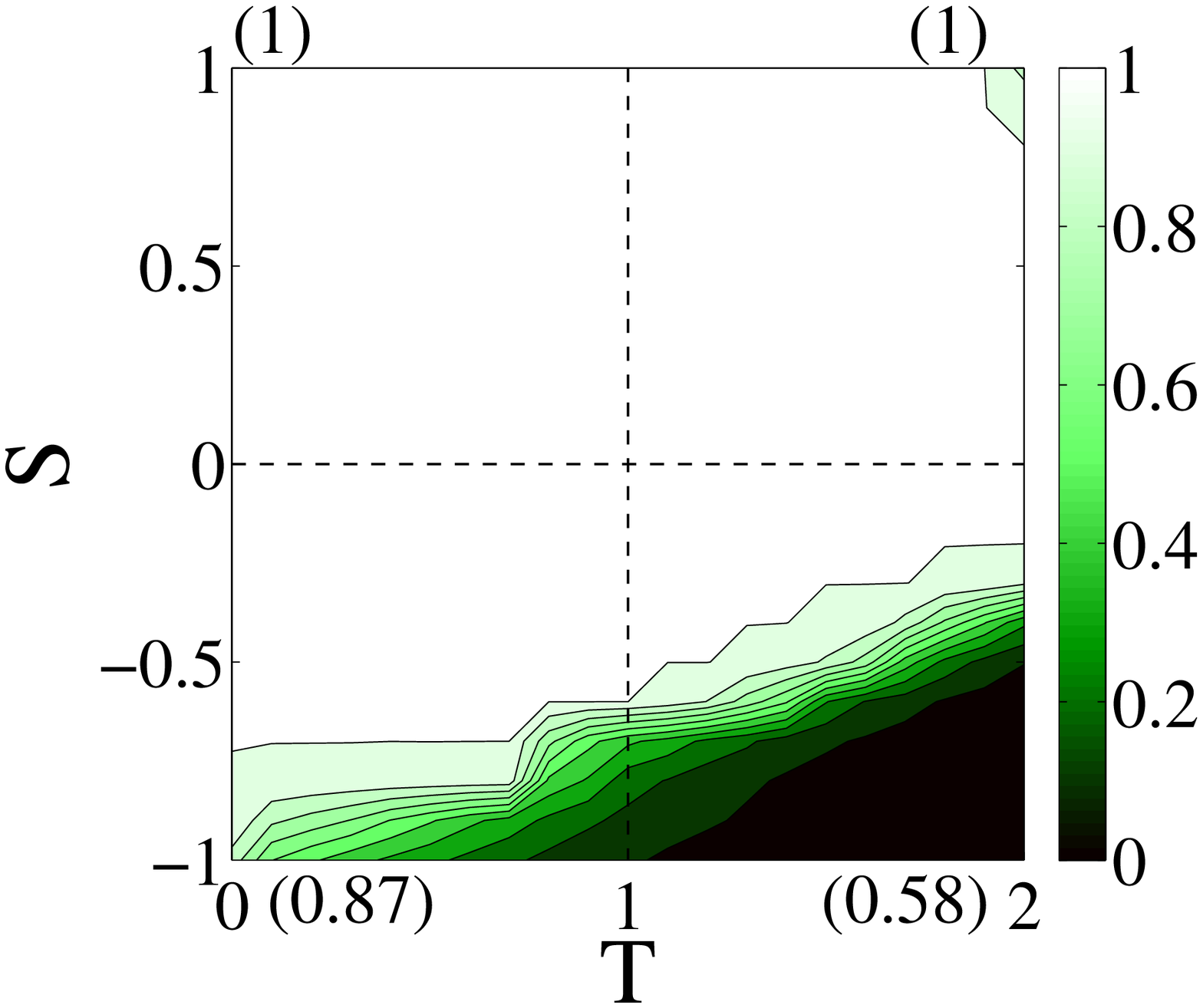} &%SuperGrid_ibr
\end{tabular}
\caption{(Color online)  Average cooperation levels on the lattice. The size of the graph is $10000$ nodes. From left to right the strategy update rules are: replicator dynamics, Fermi rule, imitation of the best, and probabilistic imitation of the
 best. In all cases the initial
 fraction of cooperators is $0.5$ randomly distributed among the graph nodes. }
 \label{latt-coop}
\end{center}
\end{figure*}

   Simple hierarchical networks
 were shown to be favorable to cooperation by using a rigorous stochastic process of the Moran type by Lieberman et al.~\cite{Lieb-Hau-Now}.
 In~\cite{weight2012} we showed how to construct relational hierarchical networks that induce high levels of cooperation.
By analogy with the latter work, in this section we construct a lattice embedded in two-dimensional space with a similar local structure and obtain high levels of cooperation. This model shows that space along with some specific constraints creates such cooperative topologies. We first place the nodes on a regular square lattice and label them according to their integer coordinates $(i,j)$. Each node with coordinates such that $i \: \mbox{mod} \: 4=r$ and $j \: \mbox{mod} \: 4=2r$ is a ``hub'' of radius $r$ which is connected to all ``small nodes'' in a square neighborhood of side $2*r+1$ and to the four closest hubs. This topology models a situation where there exist two kinds of nodes distributed in space. One kind (vertices with few connections) tries to make undirected connections to the other kind (hubs) while minimizing distances. Low-degree nodes have connections to hubs only. The hubs, in turn, form a lattice in which they are connected to the closest hubs. In Fig.~\ref{supergridimage} we show such a graph with $r= 2$. Cooperation levels  are very good in all games and for all the strategy revision rules, as seen in Fig.~\ref{latt-coop}. Indeed, the
cooperation enhancement goes beyond the best levels found in relational networks as can be seen by comparing Fig.~\ref{latt-coop} 
with  Fig.~\ref{BA}, which refers to BA networks.

\begin{figure*}[ht!]
\begin{center}
\begin{tabular} {cccccccc} 
\includegraphics[width=5.5cm]{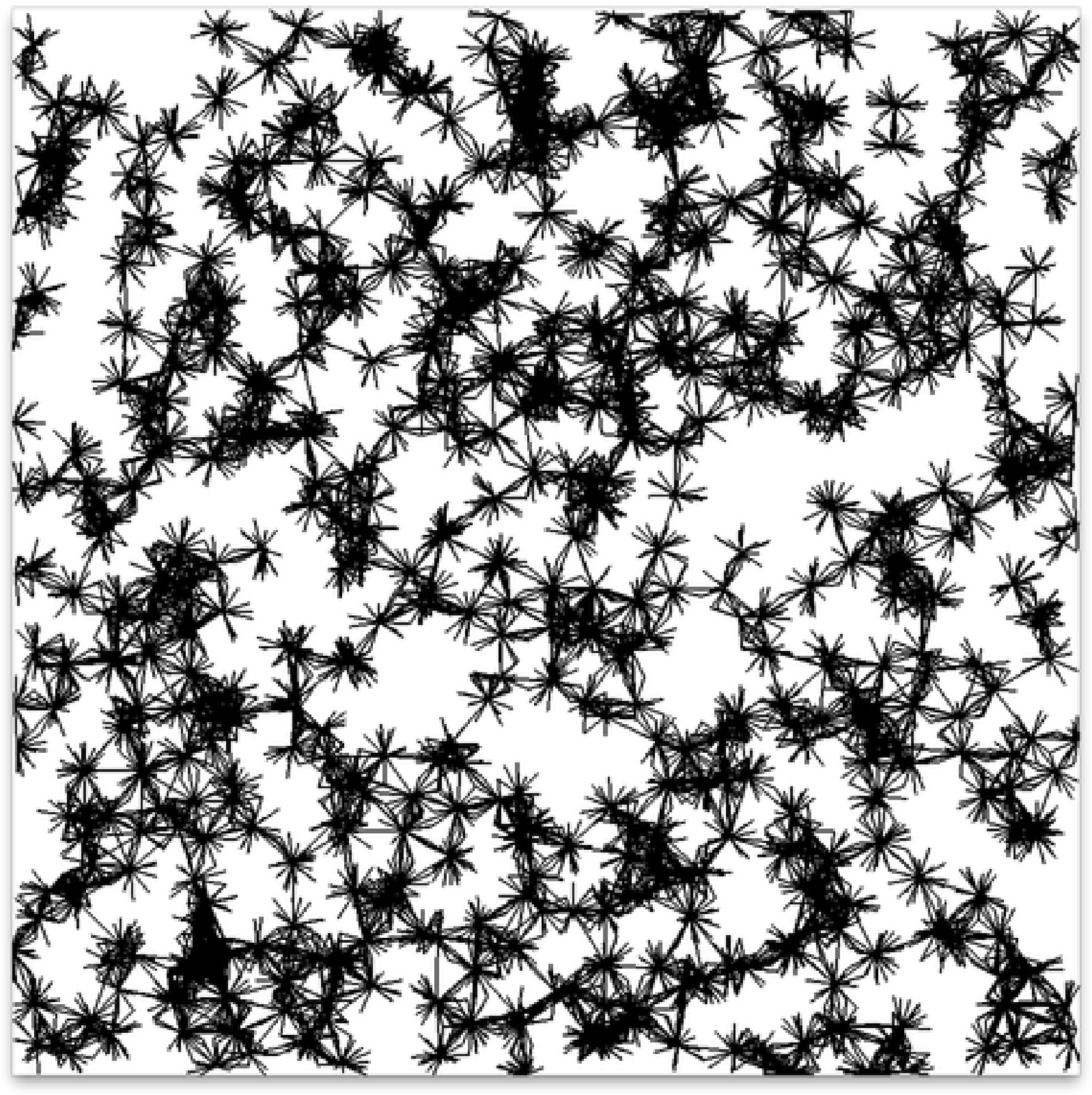} \hspace{1.3cm} &%RandomSuperGridN10000_R0p03
 \includegraphics[width=7cm]{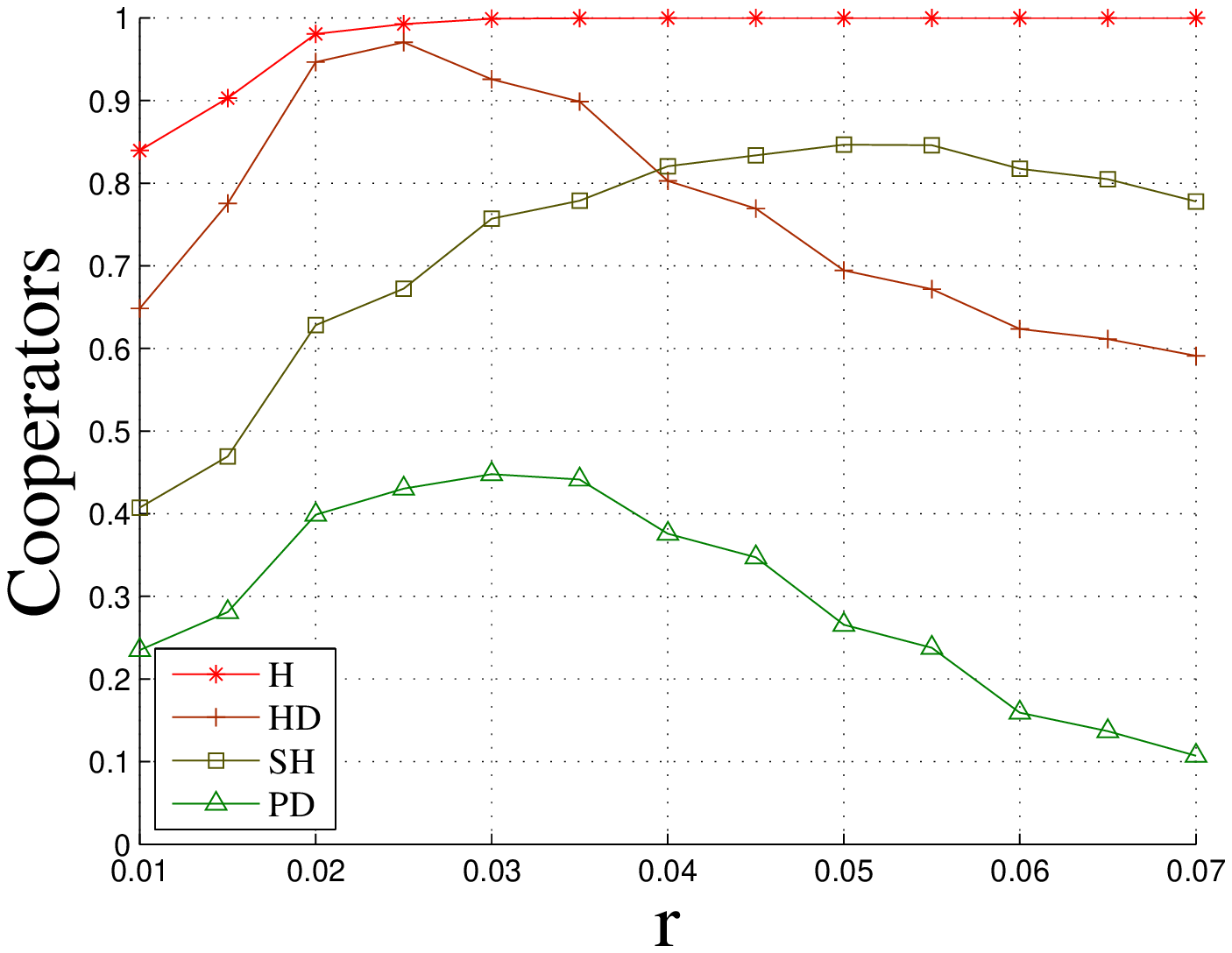}&%RandomSuperGridDiscardIsolatedVertices10000_rdBIS

\end{tabular}
\caption{(Color online)  Left : an instance of a random geometric graph with two kinds of nodes and $r=0.03$. Right: Average cooperation levels on
an ensemble of these graphs as a function of the hubs radius; the frequency of hubs is 1/16, and the radius of small vertices is null. The update rule is imitation proportional to payoff and the initial fraction of cooperators is $0.5$ randomly distributed among the graph nodes.
  In both images isolated vertices were discarded. }
 \label{randomsupergridimage}
\end{center}
\end{figure*}

Starting from random geometric graphs
with arbitrary radius distribution, in the limiting case where there are two different kinds of nodes, we show how a
network with similar properties to those of the above lattice can emerge by using the RGG model.
We constructed random geometric graphs in the same way as explained in Sect.~\ref{SFSN}, except that two nodes are linked if the sum of their radii is larger than their mutual distance. In Fig.~\ref{randomsupergridimage} (left) we used the following distribution of radius: $1/16$ of the population has an arbitrary radius $r$ and the other vertices have a null radius. The undirected resulting network is composed of hubs mainly connected to low-degree vertices, which in turn are not connected among themselves. The  low-degree vertices which are not connected to any hubs are isolated. In order to focus on the interesting part of the network, we discarded them, taking into account that they cannot change their initially attributed strategies. Fig.~\ref{randomsupergridimage} (right) shows that cooperation
is greatly enhanced for a  sizable range of radius $r$, although it doesn't reach the exceptional levels of the lattice.
The shape of the curves can be qualitatively understood noting that, when $r$ is small, say less than $0.015$, the hubs have few connections between
themselves and the network becomes fragmented into small clusters. On the other hand, when $r$ is large, the mechanism leading to cooperation 
explained in detail  
in~\cite{weight2012} doesn't work anymore. Now low-degree vertices may be connected to several hubs. This fact weakens the probability for
defector hub to imitate the strategy of a high payoff cooperator hub, since cooperator hubs are no longer surrounded by low-degree cooperators.

\subsection{Network Type and Assortativity of Strategies}
\label{assort} 

A potentially interesting question concerns the way in which strategies are distributed at steady-state among the
network nodes. At the beginning the distribution is uniform random but during the dynamics it typically
evolves and its final state could be different in different network types, according to the game played. This effect can be 
evaluated by using  several measures of ``similarity'' between vertices. Here we have chosen a measure that is inspired by Newman's
work on assortativity in networks~\cite{newman-book}~\footnote{To calculate the mean assortativity of a player at steady-state,
 we take the frequency of neighbors having the same strategy and we compute the mean over the whole network. Then we
 subtract the same quantity assuming that the strategies are randomly distributed in the same network }. 
 A state will be called ``assortative'' if cooperators tend to be
surrounded by coperators and defectors by defectors. It will be called ``disassortative'' in the opposite case, and there will
be an absence of correlations between strategies if the distribution is random. 

\begin{figure*}[ht!]
\begin{center}
\begin{tabular} {cccccccc} 
 \includegraphics[width=6cm]{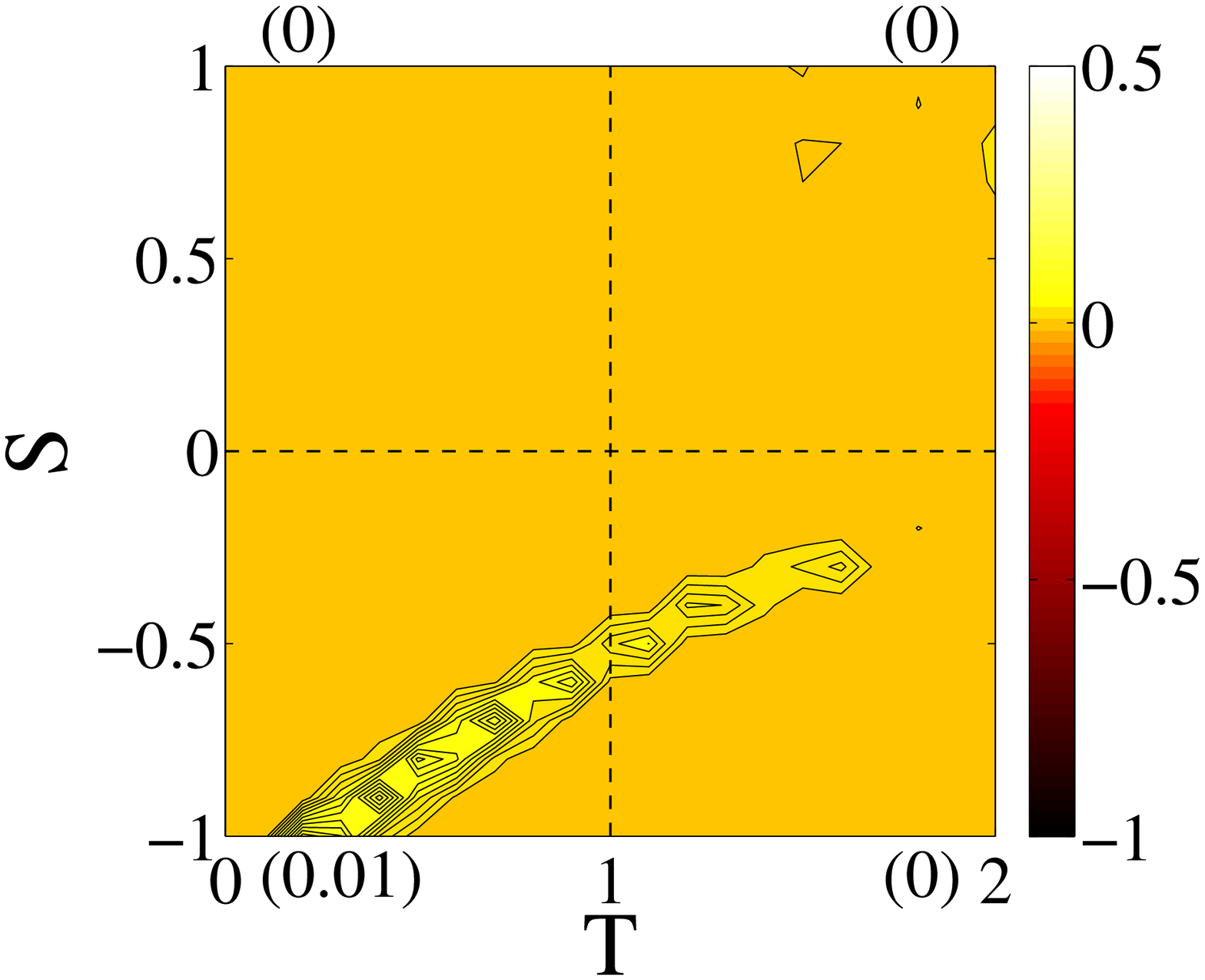} &%Apollonian_Assortativity_RD2
\includegraphics[width=6cm]{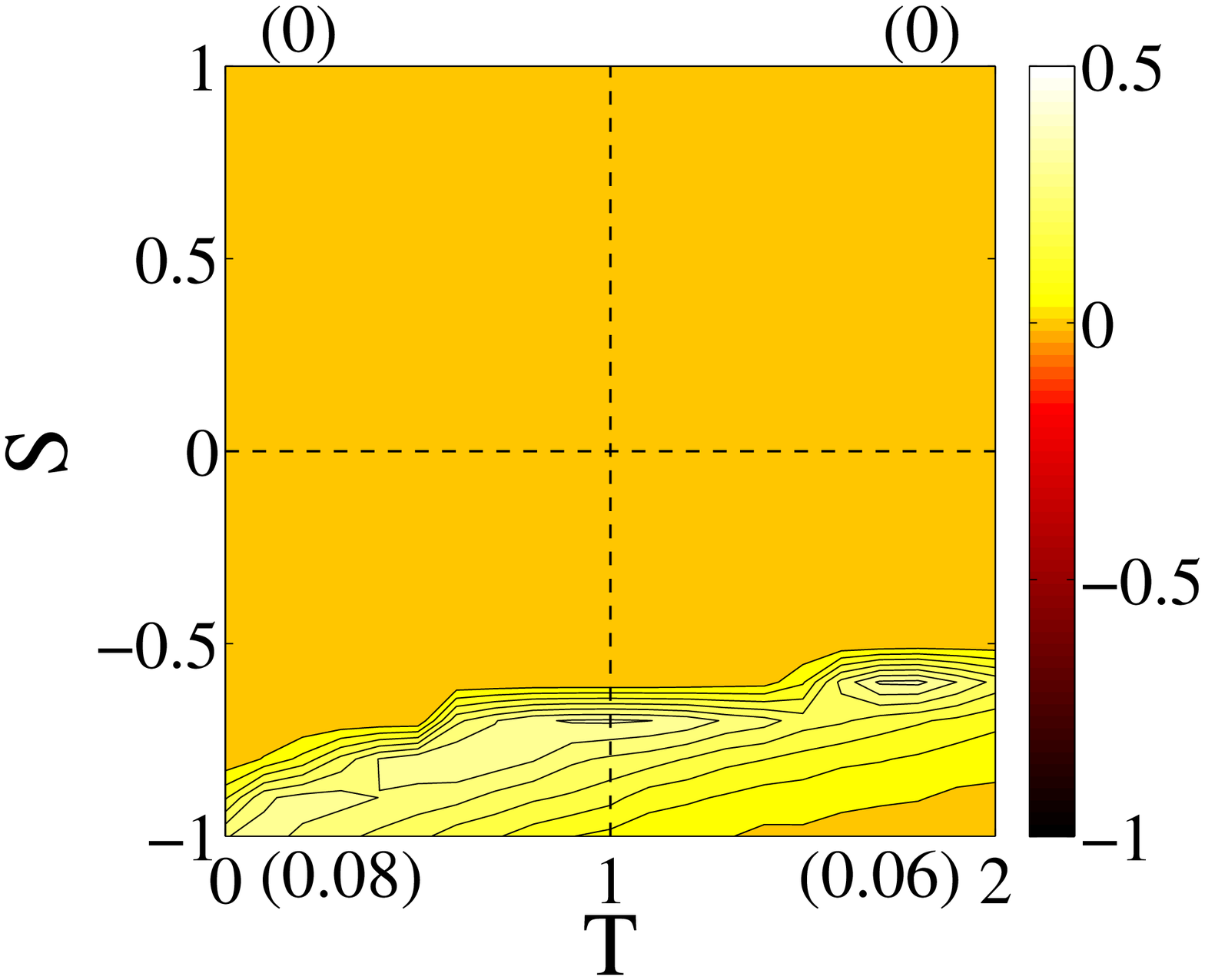} &%SuperGrid_Assortativity_RD2
\includegraphics[width=6cm]{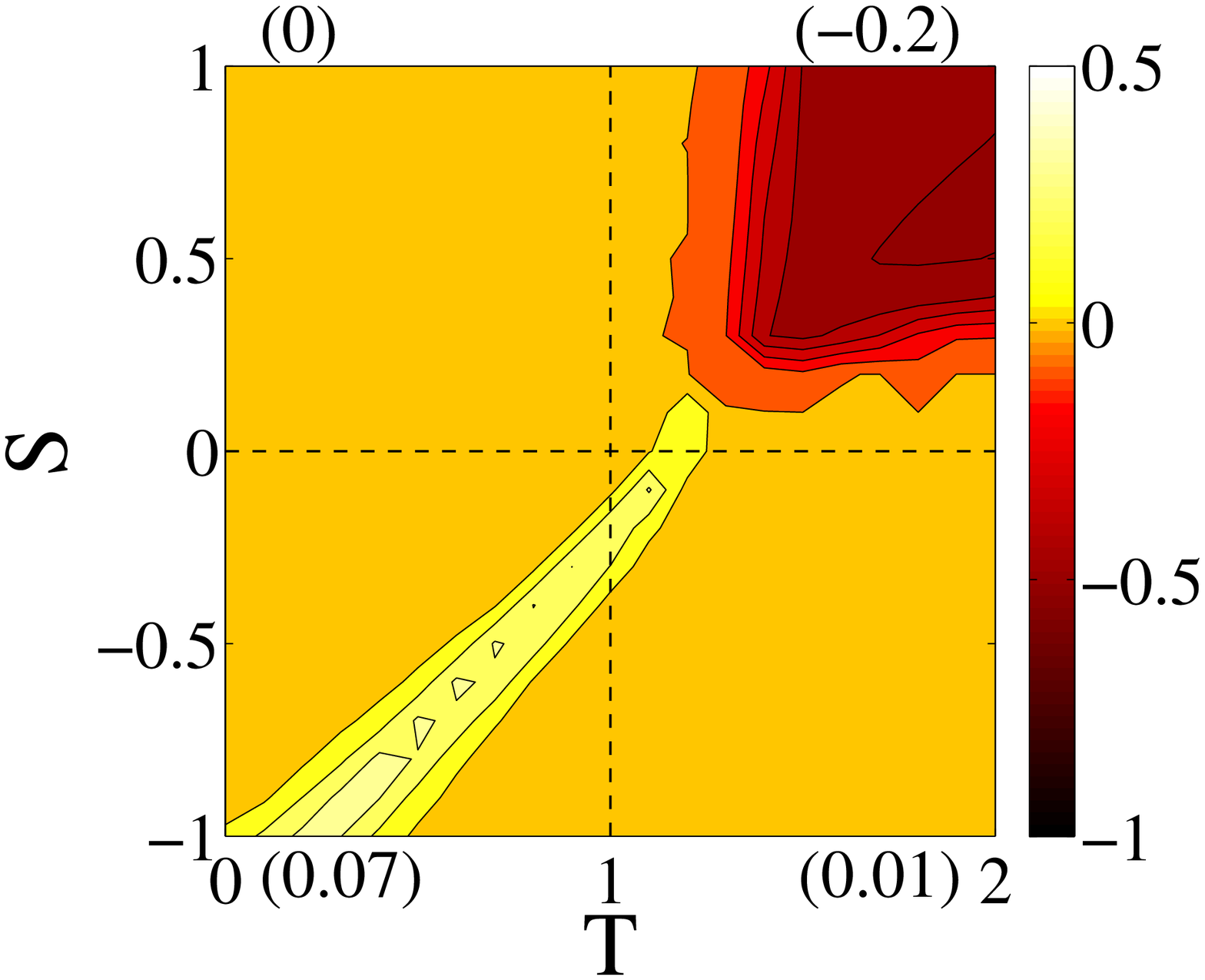} &%RGG_Assortativity_RD2
\end{tabular}
\caption{(Color online) Average strategy assortativity levels with the replicator dynamics on the Apollonian network (left image), cooperative grid (middle image), and the regular geometric graph with $\langle k \rangle  = 8$ (right image). In all cases the initial
 fraction of cooperators is $0.5$ randomly distributed among the graph nodes.}
 \label{Assort}
\end{center}
\end{figure*}

In Figs.~\ref{Assort} we report the results of our assortativity analysis for Apollonian networks (left), the spatial cooperative grid defined at the beginning of sect.~\ref{supergrid} (middle), and for RRGs (right). The first thing to notice is that strategy disassortativity is only present in the RGG for the HD quadrant, as the
HD stable equilibrium in well-mixed populations, and also partly in random graphs, consists of a mix of Cs and Ds. Now, due to the HD
payoff structure, locally a cooperator tends to be surrounded by defectors and the other way around for a defector. On the other hand, in the middle and
left images, in the HD space assortativity disappears because now the corresponding game phase space becomes totally cooperative (compare with
the leftmost images of Figs.~\ref{ApolloniusSimple} and~\ref{latt-coop}). For the SH there is an increase in cooperation too going from the RGG to
Apollonian, and especially the lattice.  The SH game features two monomorphic stable equilibria in well-mixed populations which explains why assortativity 
is near zero in most of the quadrant. Nevertheless, in the unstable middle and low region C and D can coexist thanks to the  local
network structure that allows clusters to form. But here, contrary to the HD, the strategies show
some positive assortativity since it is best for players to coordinate on the same action. In Fig.~\ref{Assort} we observe that this assortative region bends
downwards and extends to the right in the left and middle image with respect to the RGG case (right image). This region extends to the PD phase space
in the two highly cooperative networks.

\section{Discussion and Conclusions}

Evolutionary games on static spatial networks have been intensively studied in the past but mainly on two-dimensional regular lattices, either taking into account Euclidean distances explicitly or implicitly, as they can be trivially embedded in a metric space.
However,  lattices are only an approximation of actual network of contacts in geographical space. Indeed, many economic, transportation, and communication systems rely on actual positions in space and agents usually have a variable number
 of neighbors. For this reason,
here we have studied typical two-person, two-strategies evolutionary games on spatial networks having homogeneous degree
distributions such as geometric random graphs, as well as heterogeneous ones with right-skewed degree distributions
such as scale-free networks. We have studied evolutionary games on two spatial scale-free networks models: a first one based on a 
variant of Rozenfeld et al.'s construction~\cite{Rozenfeld2002} called SFSN, and the 
Apollonian networks~\cite{Apollonius2005Herrmann}. Concerning the second model, we extended previous results to a much larger parameter's space, allowing to discuss our results more accurately. We find that cooperation is promoted on spatial scale-free networks with respect to the random geometric graphs, except in the Stag-Hunt games. Now if we compare SFSNs with BA relational networks, SFSNs show the same trend but the gains are lower with all dynamics. Still, this is an interesting results since SFSNs are important in practice, for example in ad hoc 
communication networks. 
On Apollonian networks cooperation is the highest ever observed on networks of the scale-free type in all the games
studied here, although Apollonian networks  would be difficult to reproduce in practice. 
We also point out that these results are robust with respect to the several standard strategy update rules used in our
simulations.

Finally, we have introduced a two-dimensional hierarchical net whose structure has been inspired
by a previous relational graph model which highly promotes cooperation~\cite{weight2012}. On this particular spatial 
network cooperation reaches the highest
values, with $66\%$ of the population being cooperators in the average in the PD, the totality in the HD,
and $83\%$ in the SH with the local fitness-proportional rule and the Fermi rule. IB and IBR give similar figures. 
Of course, these unprecedented cooperation levels are theoretically interesting but can only be 
actually reached if the special network can be formed and kept fixed, since naturally formed networks could hardly take
this shape. 
 However, using similar ideas, we have shown that more realistic networks can be produced
that can attain a rather high level of cooperation using a modified construction starting from random geometric graphs. 

The general conclusion of this work is that promotion of cooperation in all the games' parameter space is possible
 on static networks of agents constrained to
act in geographical space, provided that agents interact according to some special spatial network of contacts
 that creates a connection hierarchy among the agents.

\section*{Acknowledgments} We thank the anonymous reviewers for their detailed and insightful remarks that
helped to improve the manuscript.
 
%\bibliographystyle{}
%\bibliography{jeux}

\end{document}